\documentclass[
 reprint,
 amsmath,amssymb,
 aps,
]{revtex4-2}

\usepackage{float}
\usepackage{sidecap}
\usepackage{color}
\usepackage{graphicx}
\usepackage{dcolumn}
\usepackage{bm}
\usepackage{hyperref}
\usepackage{ragged2e}
\usepackage{xcolor}
\hypersetup{
    colorlinks,
    linkcolor={red!50!black},
    citecolor={blue!50!black},
    urlcolor={blue!80!black}
}

\usepackage{braket}

\newtheorem{thm}{Theorem}
\newtheorem{dfn}[thm]{Definition}

\begin{document}

\preprint{APS/123-QED}

\title{
Measurement-driven navigation in many-body Hilbert space: Active-decision steering}

\author{Yaroslav Herasymenko}
\affiliation{Instituut-Lorentz, Universiteit Leiden, P.O. Box 9506, 2300 RA Leiden, The Netherlands}

\author{Igor Gornyi}
\affiliation{\mbox{Institute for Quantum Materials and Technologies, Karlsruhe Institute of Technology, 76021 Karlsruhe, Germany}}
\affiliation{\mbox{Institut f\"ur Theorie der Kondensierten Materie, Karlsruhe Institute of Technology, 76128 Karlsruhe, Germany}}
\affiliation{Ioffe Institute, 194021 St.~Petersburg, Russia}

\author{Yuval Gefen}
\affiliation{Department of Condensed Matter Physics, Weizmann Institute of Science, 7610001 Rehovot, Israel}

\begin{abstract}
The challenge of preparing a system in a designated state spans diverse facets of quantum mechanics. To complete this task of steering quantum states, one can employ quantum control through a sequence of generalized measurements which direct the system towards the target state. In an active version of this protocol, the obtained measurement readouts are used to adjust the protocol on-the-go. This enables a sped-up performance relative to the passive version of the protocol, where no active adjustments are included. In this work, we consider such active measurement-driven steering as applied to the challenging case of many-body quantum systems. The target states of highest interest would be those with multipartite entanglement.
Such state preparation in a measurement-based protocol is limited by the natural constraints for system-detector couplings. We develop a framework for finding such physically feasible couplings, based on parent Hamiltonian construction. For helpful decision-making strategies, we offer Hilbert-space-orientation techniques, comparable to those used in navigation. The first one is to tie the active-decision protocol to the greedy accumulation of the cost function, such as the target state fidelity. We show the potential of a significant speedup, employing this greedy approach to a broad family of Matrix Product State targets. For system sizes considered here, an average value of the speedup factor $f$ across this family settles about $20$, for some targets even reaching a few thousands. We also identify a subclass of Matrix Product State targets, including the ground state of the Affleck-Kennedy-Lieb-Tasaki spin chain, for which the value of $f$ increases with system size. In addition to the greedy approach, the second wayfinding technique is to map out the available measurement actions onto a Quantum State Machine. A decision-making protocol can be based on such a representation, using semiclassical heuristics. This State Machine-based approach can be applied to a more restricted set of targets, where it sometimes offers advantages over the cost function-based method. We give an example of a W-state preparation which is accelerated with this method by $f\simeq3.5$, outperforming the greedy protocol for this target.

\end{abstract}

\maketitle

\section{\label{sec:level1}Introduction\protect}

Quantum state preparation is a prominent routine in quantum information processing toolbox \cite{wang2020integrated,handel2005modelling, bloch2008quantum, kwek2012measurement, pan2001entanglement,
luo2017deterministic,
bohnet2016quantum, stockton2004deterministic, marr2003entangled, pechen2006teaching, kraus2008preparation,
shen17, iten17, piro21, lu22, stockill2017phase, shao2012engineering, lin2016preparation, liu2010deterministic}. Such procedure often implies steering a quantum system from a ``simple'' towards a more complex, pre-designated resourceful state, e.g. a many-body entangled state. A steering protocol is characterized by an as short as possible runtime and high resulting overlap with the target state.
Constructing such protocols can be done in multiple distinct ways. One is to design the Hamiltonian of the system, such that its unitary evolution leads to a designated state. This paradigm is represented by methods like digital computation or analog simulation \cite{bloch2008quantum,
pan2001entanglement,
luo2017deterministic,
bohnet2016quantum,
stockill2017phase}.
Such protocols require exact knowledge of the starting state, as well as the precise timing of the unitary evolution, to be accurate.
Another strategy is to add a dissipative element to the protocol. Combined with the Hamiltonian evolution, this results in methods such as drive-and-dissipation \cite{pechen2006teaching, kraus2008preparation} and quantum channel engineering \cite{shen17, iten17, piro21, lu22}. Finally, one can design a sequence of generalized measurements, which brings the system towards the target state via measurement back-action alone \cite{pechen2006quantum, roa2006measurement, roy2020measurement}. The system-detector coupling completely governs the relevant part of the evolution in such a protocol (see also Ref. \cite{ippo21}). Unlike protocols involving pre-defined unitary evolution, such measurement-driven state preparation may not require knowledge of the starting state and fine-tuning of the system Hamiltonian \cite{roy2020measurement}.

The above types of state-preparation strategies can be referred to as \textit{passive}, meaning that these protocols are pre-determined and pursued regardless of how the system evolves.
Given this perspective, it appears beneficial to go beyond the forms of control described above, and introduce the concept of \textit{active} decision making. This type of steering exploits information extracted during the system's evolution to decide on the operations that follow.
This is also referred to as closed-loop quantum control
and is typically used to improve the Hamiltonian-based state preparation \cite{zhang2017quantum,
lar20,
ticozzi2009analysis,
wiseman1994quantum,
belavkin2004towards,
wiseman1993quantum}; dissipation-based protocols are also being considered \cite{grigoletto2020stabilization}.
Extracting the necessary information requires introducing measurements into the protocol, which may result in an undesired back-action.
Nevertheless, in many cases, closed-loop control does yield an improvement in the speed and the fidelity of the protocol.

Another possibility, which is a subject of increasing interest, is to employ active decision in measurement-driven protocols (implying no other source of drive, Hamiltonian or dissipative) \cite{jacobs2010feedback,
ashhab2010control,
fu2014feedback}. In such protocols, information about the running state of the system is naturally available from the employed measurements. This data can be used for the active choice of subsequent generalized measurements, such that the target state is prepared as rapidly and accurately as possible \cite{
jacobs2010feedback,
ashhab2010control,
fu2014feedback}.
Some general theorems have been stated concerning such state preparation protocols \cite{fu2014feedback}, along with some specific protocols designed to reach single-qubit target states \cite{jacobs2010feedback, ashhab2010control}. However, it remains unclear how an active measurement-driven protocol can be effectively harnessed to engineer resourceful many-body states. In this case, the large size of the Hilbert space makes it challenging to actively steer the system evolution in the desired direction.

In this work, we establish a general framework for measurement-driven active navigation in Hilbert space. The goal of this framework is to construct active-decision protocols for measurement-only steering of many-body system to a given target state. 
In particular, we focus on target states manifesting genuine multipartite entanglement. The key problem here is to ensure that the protocol achieves the target state in as short time as possible. When attempting to address this problem, one is naturally constrained by a few factors. One is that only reasonably local system-detector couplings are to be used in the protocol. Moreover, we require that the number of distinct system-detectors couplings available for steering does not scale up faster than the system's size (this number should not be super extensive). This practical requirement restricts the capabilities of the protocol. From the limitations above, it naturally follows that applying one type of coupling generally leads to an update in the expected benefits from other couplings (see Sec.~\ref{sec:III} for a detailed analysis). This phenomenon, which we refer to as ``coupling frustration'', calls for nontrivial coordination between different coupling applications. Heuristically, one can view the problem at hand as one of orienteering: it is easy to ``get lost'' in the many-body Hilbert space when exploring it with a limited set of tools (cf. Ref.~\cite{lar20}).

We note that upon the availability of indefinite computational power, one can always find an optimum sequence of measurements through dynamic programming techniques (cf. Ref.~\cite{fu2014feedback}). Roughly speaking, this can be done by considering all possible future quantum trajectories of the system. However, in a large Hilbert space, it is practically impossible to realize the theoretically optimal feedback policy. This is because the extensive consideration of outcome scenarios is too complex for a many-body Hilbert space of an already not very large system (it increases at least exponentially with the system's size and the runtime of the protocol). Instead, we aim at designing heuristic strategies for active decision-making. The key metric of success is the speedup coming from such active decision making. This is defined as factor $f$ between the average runtime of a passive protocol and such of a comparable protocol that employed an active decision-making strategy (see Sec.~\ref{sec:generalities} for details). The goal of our heuristic strategies is to ensure a significant -- but not necessarily optimal -- value of the speedup factor $f$.

To meet these challenges, we introduce  \textit{Hilbert-space navigation techniques}. The first technique, which we term \textit{greedy orienteering policy}, is based on the notion of a cost function. A simple example of such a cost function is the target state infidelity. Minimizing it in a greedy protocol may already yield a significant advantage compared to the passive policy. To test this approach, we study numerically the preparation of Matrix Product State (MPS) targets. We consider uniform spin-1 MPS with bond dimension 2. For such target states generated at random, we discover that the greedy policy yields speedups $f$ up to $f_{\mathrm{max}}\sim 10^3$ and an average value about $f_{\mathrm{av}}\sim 10^1$ ($f_{\mathrm{max}}=3400$, $f_{\mathrm{av}}\simeq 19$ for target states sampled at system size $N=5$). Interestingly, among these MPS targets $f$ tends to increase with the system size for states whose parent Hamiltonian has a large enough spectral gap. This in particular holds for the Affleck-Lieb-Kennedy-Tasaki state \cite{affleck1987rigorous}, which is a well-known example from the MPS family we consider.

The second navigation technique is via mapping the Hilbert space onto a colored multigraph, referred to as the Quantum State Machine. The vertices of such a graph correspond to the basis states, and the edges represent the actions of generalized measurements. 
Upon an appropriate choice of basis states, such Quantum State Machine representations allow for improved navigation in Hilbert space. This can be done by heuristically representing it as quantum wayfinding on the graph. To substantiate this heuristic, we introduce the notion of semiclassical coarse-graining of a Quantum State Machine graph. 
Optimizing the exploration of these graphs by choosing the most appropriate system-detector couplings results in advantageous active-decision protocols. Notably, Quantum State Machine approach is conceptually different to the greedy approach, and has the potential to offer a higher speedup factor $f$ for some applications.
To exemplify this alternative navigation paradigm we consider the preparation of the 3-qubit W-state. A numerical study demonstrates that Quantum State Machine based approach yields $f=3.5$, while the greedy approach results in $f=3.1$.

Throughout the paper, we assume that we know the initial state of the system. This can be a ``cheap'' (say, product) and robust quantum state that does not require many resources for its preparation.
However, one can directly generalize the above approaches to the case where the initial state is unknown and is therefore represented by a density matrix. In the more intricate case of a Quantum State Machine-based policy, one would then need to take a weighted combination of graph navigation protocols with different initial states.

The remainder of the paper is organized as follows.
In Sec.~\ref{sec:generalities}, we introduce the basics of
measurement-induced steering. Specifically, in Sec.~\ref{sec:IIA}, we define the steering protocols and their elements, as well as the quantitative measure of the protocol's success. Then, in Sec.~\ref{sec:single_qubit}, we illustrate these definitions as applied to passive steering of a single qubit. The general selection criteria, including locality and extensivity, for the system-detector couplings, which are to be used for the active steering, are addressed in Sec.~\ref{Sec:IIc}. 
In Sec.~\ref{sec:III}, we introduce the notion of frustration of steering and discuss the possibilities of protocols' speed-up for mutually commuting (Sec.~\ref{sec:mutually_commuting}) and non-commuting (Sec.~\ref{Sec:IIIB}) couplings. In the latter case, we develop a parent-Hamiltonian approach. A ``quantum-compass'' approach to active-decision steering, based on the greedy cost-function accumulation policy, is developed in Sec.~\ref{sec:quantum_compass}. where we also employ this scheme to the preparation of the MPS states. In Sec.~\ref{sec:QSM},
we develop the Quantum State Machine framework. In Sec.~\ref{sec:QSM_generalities}, we introduce the generalities of this approach based on the underlying representation of the steering protocol in terms of a quantum graph. Next, we discuss the quantum parts of this graph (Sec.~\ref{sec:QSM_quantum_elements}), as well as the coarse-graining procedure, with the resulting coarse-grained graph being semiclassical (Sec.~\ref{sec:QSM_coarse-graining}). 
The advantage of this type of Hilbert-space orienteering is illustrated in Sec.~\ref{sec:QSM_W_preparation}, where an active-decision steering protocol for preparation of a three-qubit W-state is presented. Our findings are summarized and discussed in Sec.~\ref{sec:discussion}.

\section{Measurement-driven state preparation \label{sec:generalities}}

\subsection{Generalities \label{sec:IIA}}

Measurement-driven state steering is a specific class of state-preparation protocols. Its basic building blocks are coupling the quantum system (\textit{s}) to quantum detectors (ancillary systems) utilizing engineered interactions, followed by strong measurement on the detectors (\textit{d}). The goal of designing a measurement-based steering protocol is to generate a process that prepares the desired system state by utilizing a sequence of measurement back-actions. 

Here, we will additionally assume that the internal evolution of the system and the detector are trivial (their Hamiltonians are kept null: $H_s=0$, $H_d=0$), as in Refs.~\cite{jacobs2010feedback,
ashhab2010control,
fu2014feedback}, so that the dynamics in the problem is governed solely by the system-detector interaction $H_{s,d}$. For concreteness of analysis, we also constrain the detector to be a qubit initialized in a trivial state $\ket{0}$, and the system to be represented by $N$ spins. Although a general spin $S$ can be considered, we focus on the cases $S=1/2$ and $S=1$. We assume certain knowledge about the initial state of the system, which is described by the initial density matrix $\rho_\mathrm{in}$. For the sake of simplicity, we further address the target state which is a pure state
$\ket{\psi_{\mathrm{targ}}}$.

\begin{figure}[t]
\includegraphics[width=\linewidth]{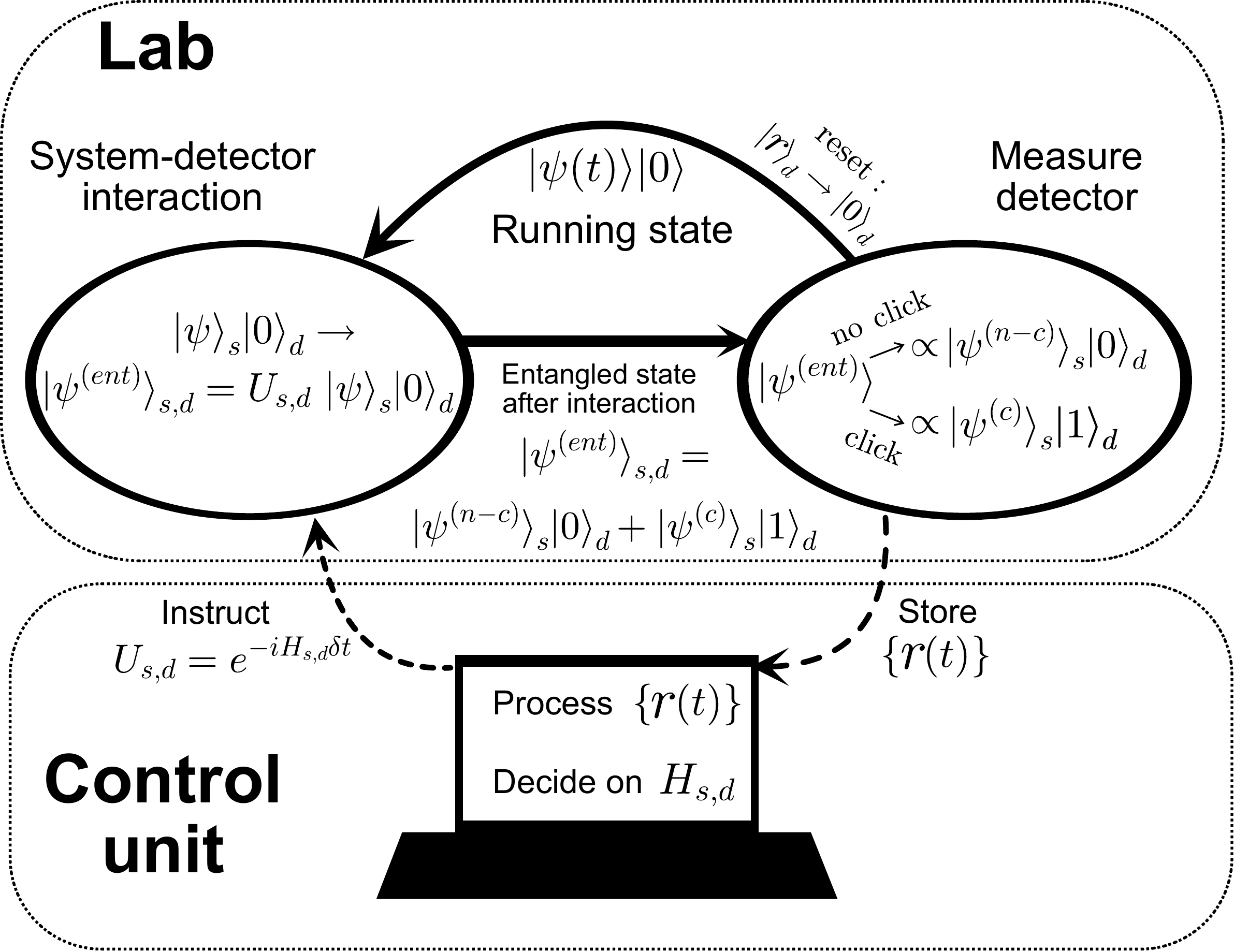}
\caption{\label{fig:MDN} Basic design of the measurement-driven state preparation. The procedure starts with a given initial state $\rho_\mathrm{in}$ and proceeds with a protocol, as described in Def.~\ref{dfn:steering_cycle}, until a good approximation of the target state $\ket{\psi_{\mathrm{targ}}}$ is achieved. The control unit decides on the system-detector interaction unitary $U_{s,d}$ based on the stored record of detector readouts. We focus on constructing an optimized policy for decision-making, such that the target state is prepared as quickly as possible. 
}
\end{figure}

Although the ensuing protocol can be further generalized (see Sec.~\ref{sec:discussion}), we now formally fix its structure as given below: 

\begin{dfn}
\label{dfn:steering_cycle}

A measurement-driven state steering is a protocol that is performed to prepare a state $\ket{\psi_{\mathrm{targ}}}$, starting from the state $\rho_\mathrm{in}$. It runs by repeating iterative cycles of the following form (see Fig.~\ref{fig:MDN}):

\begin{enumerate}
    \item \label{preparation} Prepare the detector qubit in the state $|0\rangle$.
    \item \label{decision} Based on the available information, select the system-detector coupling Hamiltonian $H_{s, d}$ to be used in the next step. 
    \item \label{interaction} Perform a system-detector evolution governed by a Hamiltonian $H_{s, d}$ for a short time interval $\delta t$: $U_{s,d}=e^{-iH_{s,d}\delta t}$. 
    \item \label{measurement} Once the system-detector evolution is over, projectively measure the detector qubit in the $Z$-basis. Store the readout $r$ for further processing.
    \item \label{termination} Decide whether the protocol is to be continued or terminated. In the former case, return to step 1.
    
\end{enumerate}

\end{dfn}
Now, in the vast space of protocols that have such structure, we would like to emphasize the distinction between two classes of protocols: passive and active. 

In a passive protocol, the stored readouts $\{r(t)\}$ from step \ref{measurement} may influence the decision for protocol termination or continuation at step \ref{termination}, but not the choice for the interaction Hamiltonian $H_{s,d}$ made at step \ref{decision} in the next protocol cycles.
Hamiltonians $H_{s,d}$ can still be chosen differently for different iterations: e.g. for a large system, the detector qubit can be coupled to different subsystems thereof. However, $H_{s,d}$ used at each cycle in the passive protocol has to be pre-determined from the outset. If a passive protocol also has a pre-determined duration (and thus doesn't use readouts $\{r(t)\}$ at the termination step), we land in a subclass of passive protocols where the readouts don't have any influence on the protocol. We would refer to such protocols as ``blind steering''. For blind steering, the readouts of the detector at any given cycle can be averaged, i.e., following the measurement, the detector's density matrix is traced out. In this work, however, we will focus on the non-blind version of passive steering, where readouts are indeed employed for an informed protocol termination.

In contrast to passive protocols, in an active protocol one uses the readouts $\{r(t)\}$ to make an informed decision for the interaction Hamiltonians $H_{s,d}$ as well as for termination/continuation of the protocol.
Active decision-making has to follow a certain policy, which becomes the crucial part of the protocol. For a good active policy, its adoption should result in a significant speedup of the protocol compared to its passive counterpart. 
Alternatively, one can also fix the protocol runtime and aim to improve the precision of the state preparation.
We focus on the former target: minimizing protocol runtime for a fixed target precision. The major challenge in this work is to construct such advantageous active decision-making policies. By comparing active steering with the (non-blind) passive steering as defined above, we investigate the advantage offered specifically by the directed evolution, i.e. active decision-making for $H_{s,d}$.

Before we move on to the issue of active policy constructions, let us discuss the criteria for termination of a running protocol. 
In general, one cannot guarantee ``perfect steering,'' i.e., obtaining the desired target state with the fidelity of $1$ in a finite number of protocol cycles. Instead, one may consider preparing the target state with \textit{infidelity} $R$:
\begin{equation}
R\left(\rho_\mathrm{out}, \ket{\psi_{\mathrm{targ}}}\right)
\equiv 1-\langle\psi_{\mathrm{targ}}|\rho_\mathrm{out}|\psi_{\mathrm{targ}}\rangle,
\label{eq:infidelity}
\end{equation}
where the state $\rho_\mathrm{out}$ is the final state of the system once the protocol is terminated. It is worth emphasizing that the system evolution during the protocol is probabilistic and depends on the stochastic readouts $\{r(t)\}$. It follows that different runs of the same protocol may yield different $\rho_\mathrm{out}$ and, thus, the infidelity R. 
Therefore, to characterize the protocol as a whole, we introduce the following accuracy measure:

\begin{dfn}
\label{dfn:epsilon_precise}

We refer to a measurement-driven state-preparation protocol as $\epsilon$-precise, if the infidelity between the final state and the target state is bounded by $\epsilon$ for any run of the protocol:
\begin{equation}
R\left(\rho_\mathrm{out}, \ket{\psi_{\mathrm{targ}}}\right)
<\epsilon.
\label{eq:epsilon_precise}
\end{equation}
\end{dfn}

Given the knowledge of the readout sequence, we may simulate the quantum system state (the quantum trajectory) on a computer in parallel to the measurement run. Thus one can infer the running system state exactly (referred to as filtering in the literature \cite{zhang2017quantum}), and test inequality \eqref{eq:epsilon_precise}. This sets a trivial criterion for protocol termination, which we will apply by default to all passive and active protocols considered in this work. Namely, a protocol can be terminated right after the cycle when the target state infidelity becomes smaller than $\epsilon$, thus making it an $\epsilon$-precise protocol. Apart from controlling the precision, we are interested in the number of cycles $\tau$ after which the protocol has been terminated. As $\tau$ may differ greatly, depending on a specific run, we will characterize the protocol by the runtime averaged over many runs
$\tau_\mathrm{av}\equiv\langle \tau \rangle_{\mathrm{run}}$. Here averaging is taken over stochastic readout sequences. In reality, steering errors as well as external noise may be further contributing factors to the stochasticity.
For the given target state and target precision $\epsilon$ 
(cf. Def.~\ref{dfn:epsilon_precise}), our goal is to find an $\epsilon$-precise protocol such that $\tau_\mathrm{av}$ is as small as possible. We will be considering this minimization as the key goal of our constructions. In particular, when developing the active policy, we will be interested in maximizing the speedup factor
\begin{equation}
    f\equiv\frac{\tau^\mathrm{(pas)}_\mathrm{av}}{\tau^\mathrm{(act)}_\mathrm{av}}
\end{equation} 
of the active protocol relative to its passive counterpart.

\subsection{Passive steering: Single qubit \label{sec:single_qubit}}

As a simple example of a measurement-driven protocol, we consider single-qubit steering (for a more general consideration, the reader is referred to Sec.~\ref{Sec:IIc}). For simplicity, we will assume the target state to be $|0\rangle$, and the starting state to be a perfectly mixed state: $\rho_\mathrm{in}=\mathrm{diag}(1/2,1/2)$. A single coupling suffices to guarantee the preparation of the target state (in fact, from an arbitrary starting state) with an arbitrary precision \cite{roy2020measurement}:
\begin{equation}
H_{s,d} = \gamma \sigma_s^-\sigma_{d}^+ + \text{H.c.}
    \label{eq:single_qubit_coupling}
\end{equation}
Here, $\sigma_s$ and $\sigma_{d}$ are Pauli matrices acting in the system and detector spaces, respectively. 
By construction, a protocol that operates with 
only a single coupling Hamiltonian $H_{s,d}$, i.e., without a readout-based option of choosing different couplings, is considered passive. Nevertheless,
even for passive protocols, one can introduce a policy based on the measurement outcomes, which would accelerate quantum-state steering. 

Let us first address a protocol that runs for $\tau$ cycles using the coupling \eqref{eq:single_qubit_coupling}, regardless of the measurement outcomes. Under the definition given in Sec.~\ref{sec:IIA}, this would be an example of \textit{blind} steering. In this case, the probability of obtaining a readout $r=0$ decreases exponentially with the total number of cycles $\tau$. 
Tracing out detector outcomes (since we are blind to measurement outcomes), this results in a density matrix:
\begin{equation}
     \rho(\tau)=
     \begin{pmatrix}
     1-e^{-\tau\gamma^2 \delta t^2}/2 & 0\\
     0 & e^{-\tau\gamma^2 \delta t^2}/2
     \end{pmatrix}.
 \end{equation}
Given the threshold infidelity $\epsilon$, we need to run the protocol for $\tau^\text{(blind)}(\epsilon)$ cycles: 
 \begin{equation}
 \tau^\text{(blind)} (\epsilon)=\frac{1}{\gamma^2 \delta t^2} \log \left(\frac{1}{2\epsilon}\right)
 \label{eq:single_qubit_passive_runtime}
 \end{equation}
This characterizes the efficiency of the completely blind passive protocol \cite{roy2020measurement} for the single-qubit setup.   

Next, we consider passive protocols where the sequence of readouts is recorded. One then needs to interpret the measurement outcomes, which for this setup is straightforward.
We note that when the readout is $r=1$ (click event), the target state is instantly prepared 
[cf. Eq.~\eqref{eq:click_scenario_map}].
Therefore, one can terminate the protocol directly after the detector clicks for the first time: in this case, all further cycles are simply redundant and do not result in any evolution of the system. 
This will constitute a termination-policy improvement of the passive blind protocol for this single-qubit case. 
If $r=0$, i.e. no click is measured (such a null-measurement \cite{zilberberg2013null} event still gives the system a nudge towards the target state by measurement back-action), the protocol simply continues until a certain maximal number of cycles, $\tau_\mathrm{max}$. The target infidelity $\epsilon$ would be directly related to $\tau_\mathrm{max}$ in a way equivalent to the blind protocol runtime \eqref{eq:single_qubit_passive_runtime}.
The average runtime of the non-blind passive protocol is then given by:
\begin{equation}
 \tau^\text{(pass)}_\mathrm{av} = 
 \frac{1}{2 \gamma^{2}\delta t^{2}} \left(1-e^{-\gamma^2\delta t^2 \tau_\mathrm{max}}\right)+
 \frac{\tau_\mathrm{max}}{2}.
 \label{eq:single_qubit_active_runtime}
 \end{equation} 
This runtime is strictly smaller than the runtime for the passive blind protocol, Eq.~\eqref{eq:single_qubit_passive_runtime}, and yields a twofold speedup in the $\epsilon\rightarrow 0$ limit. 

For a many-body system, the termination policy will not exhaust the possible active policies as it does for a single-qubit target. Indeed, a single click of the detector coupled to a subsystem does not guarantee that the whole system
is steered to the desired state. Nevertheless,
the above simple example shows that detector readouts 
can be used for accelerating the state preparation. In what follows, we will focus on active feedback strategies. There, instead of protocol termination, the local-measurement outcomes are employed for choosing the most efficient sequence of further measurement cycles.

\subsection{Selection criteria for system-detector couplings \label{Sec:IIc}}

Both for the active and passive protocols, a key feature is the choice of coupling Hamiltonians $H_{s,d}$. Given the target state $|\psi_{\mathrm{targ}}\rangle$, it is natural to constrain this choice to a certain family $\{H_{s,d}(\textbf{p})\}$, for a set of parameters $\textbf{p}$. A passive protocol can then take a periodic structure: couplings $\{H_{s,d}(\textbf{p})\}$ are employed in a predefined order, and this sequence of protocol cycles is repeated once the list of couplings has been exhausted. In an active protocol, the choice of $H_{s,d}$ in each protocol cycle translates into actively selecting the value of $\textbf{p}$. Before discussing the policies for doing so, we will consider a more basic question -- how should the family $\{H_{s,d}(\textbf{p})\}$ be constructed to yield a viable \textit{passive} protocol?

\subsubsection{Single cycle scenarios}

To understand the performance of the protocol defined by the family $\{H_{s,d}(\textbf{p})\}$, we first analyze the change of the system state after a single protocol cycle (Def.~\ref{dfn:steering_cycle}). To do so, we consider a general decomposition
\begin{equation}
H_{s,d} = V_s\sigma_{d}^+ + V_s^\dag\sigma_{d}^- + \tilde{V}_s\sigma^z_d.
\label{eq:H_s-d_decomposition}              
\end{equation}
Here $V_s$ and $\tilde{V}_s$ are arbitrary (not necessarily hermitian) system operators and matrices $\sigma^{\pm}_d=\frac{1}{2}(\sigma^{x}_d\pm i\sigma^{y}_d)$ act on the detector. 
In Eq.~\eqref{eq:H_s-d_decomposition}, we discard any terms of the form $\sim \mathbb{I}_d$, as those represent the internal system evolution. To simplify our further considerations, we will also impose a constraint $\tilde{V}_s=0$. 

With Eq.~\eqref{eq:H_s-d_decomposition} in mind, we now consider the effect of a single protocol cycle on the system state $\rho$ when the measurement outcomes are averaged over (blind measurement). 
In the weak measurement limit, $\delta t \rightarrow 0$, this is represented by the map:
\begin{align}
    \rho &\rightarrow \Lambda_{V_s}(\rho)\notag
    \\
    &\equiv \left(1-\frac{\delta t^2}{2} V_s^{\dag} V_s\right)\rho\left(1-\frac{\delta t^2}{2} V_s^{\dag} V_s\right) + V_s \rho V_s^\dag \delta t^2.
    \label{eq:steering_map}
\end{align}
The terms of order $\mathcal{O}(\delta t^2)$ in this expression represent the standard Lindbladian jump operator. 
Based on the map \eqref{eq:steering_map} for infinitesimal $\delta t$, 
one can derive a Lindblad equation describing the system 
evolution for the blind steering \cite{roy2020measurement}.

Let us now turn to the non-blind protocol, where the different measurement outcomes are discriminated. During step \ref{measurement} of the protocol cycle (cf. Def.~\ref{dfn:steering_cycle}), 
there is a probability 
$$ p^\text{(cl)}\left(\rho, V_s\right)=\delta t^2 \, \mathrm{tr}(V_s \rho V_s^\dag)$$ 
that a qubit flip is measured in the detector (click probability). 
The resulting state in the limit of small $\delta t$ is then
\begin{equation}
    \rho\rightarrow \Lambda^\text{(cl)}_{V_s}(\rho) 
    \equiv \frac{V_s \rho V_s^\dag}{ \mathrm{tr}(V_s \rho V_s^\dag)}.
    \label{eq:click_scenario_map}
\end{equation}
A ``no-click'' scenario occurs with probability $$p^\text{(ncl)}
\left(\rho, V_s\right)=1-p^\text{(cl)}\left(\rho, V_s\right),$$ 
and results in a state
\begin{equation}
    \rho \rightarrow \Lambda^\text{(ncl)}_{V_s}(\rho)
    \equiv\frac{\left(1-\frac{\delta t^2}{2} V_s^{\dag} V_s\right)\!\rho\!\left(1-\frac{\delta t^2}{2} V_s^{\dag} V_s\right)}{1-\delta t^2\mathrm{tr}(V_s^{\dag} V_s \rho)}.
    \label{eq:no-click_scenario_map}
\end{equation}
For the weak-measurement limit considered here ($||V_s\delta t||\ll 1$), the click probability is parametrically smaller than that for the no-click event: a qubit flip can be recorded in the detector only rarely.

\subsubsection{Necessary conditions for the coupling operators \label{sec:locality}}

We can now expound our considerations for the family $\{H_{s,d}(\textbf{p})\}$ in terms of the operators $\{V_s(\textbf{p})\}$. 
For a meaningful comparison between active and passive protocols, we first require that there exists a passive protocol that employs Hamiltonians $\{H_{s,d}(\textbf{p})\}$ to reach the target state $|\psi_{\mathrm{targ}}\rangle$. To ensure this convergence of the passive protocol, it is natural to demand that none of $\{H_{s,d}(\textbf{p})\}$ can move the system state away from the target state. 
Given Eqs.~\eqref{eq:click_scenario_map} and \eqref{eq:no-click_scenario_map}, this yields a 
dark-state condition $V_s(\mathbf{p})|\psi_{\mathrm{targ}}\rangle=0$ for every $\mathbf{p}$. 
This is equivalent to every operator $V_s(\mathbf{p})$ taking the form
\begin{equation}
\label{eq:target_state_annihilator}
    V_s=\sum^{D-1}_{\alpha=1} v_\alpha |\psi_{\mathrm{targ}}\rangle\langle\psi_\alpha| + \sum^{D-1}_{\alpha,\beta=1} w_{\alpha\beta} |\psi_\beta\rangle\langle\psi_\alpha|,
\end{equation}
where $D$ is the Hilbert-space dimensionality of the system, and $\{|\psi_{\mathrm{targ}}\rangle\}\cup\{|\psi_\alpha\rangle\}$ is any basis of the system's Hilbert space that includes $|\psi_{\mathrm{targ}}\rangle$ as a basis state. 

Having in mind steering of many-body states, we further require that the couplings $\{V_s(\textbf{p})\}$ are feasible to engineer in an experimental realization of the system.  In this work, we focus on the most basic aspect of this condition: locality. 
One may consider two types of locality: geometric and operator ($k$-locality \cite{kitaev2002classical}). Geometric locality of the operator $V_s$ implies that such interaction only requires coupling the system spins that are in geometrical proximity during the experiment. A $k$-local operator $V_s$ implies that only $k$ system spins are coupled at a time. 
It is natural to impose the locality constraint not on the full operator $V_s$, but its individual terms. For example, if $V_s$ involves all system spins, but its individual terms only couple $2$ spins at a time, we will consider $V_s$ a $2$-local coupling (in line with \cite{kitaev2002classical}). A $k$-local operator $V_s$ implies an interaction Hamiltonian $H_{s,d}$ that is $(k+1)$-local.

\subsubsection{Possibility of spurious dark states. \\Room for active decision-making}
It is worth stressing that $V_s(\mathbf{p})$ following the form given by 
Eq.~\eqref{eq:target_state_annihilator} for all $\mathbf{p}$ is necessary but not sufficient for $|\psi_{\mathrm{targ}}\rangle$ to be the \textit{only} dark state of the passive protocol. 
For some choices of such a family $\{V_s(\mathbf{p})\}$, a spurious final state $|\psi'_\mathrm{targ}\rangle \neq |\psi_{\mathrm{targ}}\rangle$ might be reached. However, this would imply a dark-state condition $V_s(\mathbf{p})|\psi'_\mathrm{targ}\rangle=0$ (for every $\mathbf{p}$), and this should not hold even for a single \textit{generic} operator $V_s$, which satisfies $V_s|\psi_{\mathrm{targ}}\rangle=0$. In other words, one does not expect $|\psi'_\mathrm{targ}\rangle$ to exist for generic (say, random-matrix-type) coefficients $v_\alpha$ and $\omega_{\alpha\beta}$ in \eqref{eq:target_state_annihilator}. For that, an extra constraint is needed, such as vanishing of certain $v_\alpha$, $\omega_{\alpha\beta}$, or a specific relation between the coefficients. 

One concludes that a family consisting of a single Eq.~\eqref{eq:target_state_annihilator}-type coupling $V_s$ is sufficient to prepare $\ket{\psi_{\mathrm{targ}}}$ in a passive protocol without generating a dark space. 
Notably, reducing the family to a single member would leave no room for active decision-making in a protocol defined by this family (an active protocol requires at least two operators to choose from). On the other hand, such an ultimate $V_s$ would not in general satisfy the locality conditions and, thus, would be unrealistic to implement. Natural counterexample couplings $V'_s$ that have multiple dark states arise in the important case when $V'_s$ acts only on a part of the system. 

To construct such a counterexample, one may start from an arbitrary operator $V_s$ that satisfies the dark-state condition $V_s \ket{\psi_{\mathrm{targ}}}=0$ for a single state $\ket{\psi_{\mathrm{targ}}}$ in a given system. 
Now, consider a larger system embedding the original one and construct a different target state which is a tensor product of $\ket{\psi_{\mathrm{targ}}}$ and a certain auxiliary state: $\ket{\Psi_{\mathrm{targ}}}\equiv |\psi_{\mathrm{targ}}\rangle\otimes |\tilde{\psi}_{\mathrm{targ}}\rangle$. In this case, one may take $V_s \rightarrow V'_s$,
where $V'_s = V_s \otimes \mathbb{I}_{\tilde{s}}$
still satisfies condition $V'_s \ket{\Psi_{\mathrm{targ}}}=0$ relative to this new target state in the extended Hilbert space. Yet for a general starting state of the total system, the operator $V'_s$ is obviously not sufficient to prepare the extended target state $\ket{\Psi_{\mathrm{targ}}}$. This implies the existence of spurious dark states: in fact, all states of the form $\ket{\psi_{\mathrm{targ}}}\otimes\ket{\tilde{\psi}_\mathrm{targ}}$ turn out to be dark for arbitrary $\ket{\tilde{\psi}_\mathrm{targ}}$. This links to the previous discussion: an operator $V_s$ capable of steering to a unique dark state is expected to be highly nonlocal in general, in contrast to the limited capacity of a localized operator $V_s \otimes \mathbb{I}_{\tilde{s}}$. 

From the arguments above, we conclude that a family of \textit{multiple} operators $\{V_s(\textbf{p})\}$ is needed to realistically prepare a target state, once that state is sufficiently complicated. Having multiple operators in $\{V_s(\textbf{p})\}$, in turn, opens the door for gaining advantage through active decision-making.

\section{Types of system-detector couplings  \label{sec:III}}

The preselected family of coupling operators $\{V_s(\mathbf{p})\}$ determines both the performance of the ensuing passive protocols and the possibilities for active policy construction. In the present section, we identify the crucial role of the commutation properties of $\{V_s(\mathbf{p})\}$. We first consider $N$-qubit steering protocols which employ coupling operators $\{V_s(\mathbf{p})\}$ that are mutually commuting. As a shorthand, we call this frustration-free steering. We show that a realistic passive protocol of this type can be designed for product states and certain graph states. Commuting couplings also allow for a simple feedback strategy, which results in a significant speedup of the respective passive protocol. Next, we move on to passive steering protocols that are frustrated. Such frustration of local couplings naturally arises for many-body target states related to local parent Hamiltonians. We propose an explicit method of constructing a family of non-commuting operators $\{V_s(\mathbf{p})\}$ that allows to prepare such a many-body target state in a passive protocol. This forms the basis for Sec.~\ref{sec:quantum_compass} and \ref{sec:QSM}, where we move on to the active versions of frustrated steering protocols.

\subsection{Mutually commuting couplings
\label{sec:mutually_commuting}}

Here we focus on $N$-qubit steering protocols implemented with mutually commuting couplings $\{V_s(\mathbf{p})\}$. As will be demonstrated, a passive protocol of this type can be constructed for an arbitrary target state, yielding an asymptotically precise passive preparation. For a general target this construction yields \textit{non-local} couplings $\{V_s(\mathbf{p})\}$, which are therefore impractical. We identify an exception to this rule -- a subclass of graph states that can be obtained using \textit{local} commuting couplings. For this, we discuss the constraints coming from both geometric locality, as well as $k$-locality. Finally, we extend the discussion from such passive protocols to their active counterparts. To achieve this, we propose a simple feedback strategy that speeds up frustration-free steering in a substantial way.

As a trivial example of frustration-free steering, consider an $N$-qubit product state as a target state, e.g., $|00..0\rangle$. The starting state will be assumed to be the perfectly mixed state. To prepare the target with a steering protocol, one can use a set of couplings parameterized by the qubit number $i=1,..N$:
\begin{equation}
    V^\text{(prod)}_s(i)= \gamma \sigma^-_i.
    \label{eq:product_state_couplings}
\end{equation}
Passively alternating between the steering cycles employing $V^\text{(prod)}_s(i)$ with different $i$ guarantees preparation of the target state with any given accuracy. This directly follows from the analysis of Sec.~\ref{sec:single_qubit} and \ref{Sec:IIc}. For an active version of the protocol, partial protocol termination can be applied: if a click is registered when measuring any qubit $i$, the coupling $V^\text{(prod)}_s(i)$ is dropped out from the sequence of couplings that will be applied in further cycles.
In other words, the steering with this ``fired'' coupling is terminated at this point, whereas other couplings remain active -- hence the term ``partial termination''. Since this implies a readout-based decision on the set of steering couplings that are used at a given cycle, we classify this as an active steering protocol. For the perfectly mixed $\rho_\mathrm{in}$ in the high precision limit $\epsilon\rightarrow 0$, this strategy results in the following relation between active and passive runtimes:
\begin{equation}
    \tau^\text{(act)}_\mathrm{av}(\epsilon)\simeq \frac{\tau^\text{(pass)}_\mathrm{av}(\epsilon)}{2}+
    \frac{N}{2\gamma^2 \delta t^2},
    \label{eq:multi_qubit_non_frustrated_active_runtime}
\end{equation}
which leads to a $f\simeq 2$ speedup in the limit of small $\epsilon$ -- similarly to Eq.~\eqref{eq:single_qubit_active_runtime}. 

Frustration-free steering towards any target state $\ket{\psi_{\mathrm{targ}}}$ can in principle be designed if we allow for an arbitrary coupling set. For that, consider a many-body unitary transformation to $\ket{\psi_{\mathrm{targ}}}$ from a product state $\ket{00..0}$, i.e., $\ket{\psi_{\mathrm{targ}}}=U_{\psi}\ket{00..0}$ (dropping subscript $\vphantom{\psi}_\mathrm{targ}$ for brevity). One may then formally construct a family of couplings:
\begin{equation}
    V^{(U_{\psi})}_s(i)= \gamma U_{\psi} \sigma^-_i U^\dag_{\psi}.
    \label{eq:U_psi_couplings}
\end{equation}
Any protocol for $\ket{\psi_{\mathrm{targ}}}$ preparation using couplings of the form of Eq.~\eqref{eq:U_psi_couplings} would be a unitary equivalent of the same protocol which uses couplings of Eq.~\eqref{eq:product_state_couplings} to prepare $\ket{00..0}$. Therefore, a passive protocol iterating over $V^{(U_{\psi})}_s(i)$ for different $i$ would successfully prepare the target state $\ket{\psi_{\mathrm{targ}}}$. We also conclude that a partial-termination strategy can be applied to this coupling set with the same effect as for the product state target. 
Note, however, that in most cases employing $V^{(U_{\psi})}_s(i)$ would not be practically feasible. Indeed, since $U_{\psi}$ is a general many-body operation, the couplings $V^{(U_{\psi})}_s(i)$ would involve arbitrarily non-local terms.
For most $N$-qubit states $\ket{\psi_{\mathrm{targ}}}$ with large $N$, one thus expects that the resulting $V^{(U_{\psi})}_s(i)$ would break any requirement of geometric or $k$-locality.

This locality-violation rule can be circumvented for $U_{\psi}$ which is given by a shallow circuit and thus $\ket{\psi_{\mathrm{targ}}}$ which is weakly entangled. As a resourceful example of such $\ket{\psi_{\mathrm{targ}}}$, consider a graph state defined on a generic graph $G$ \cite{hein2004multiparty}:
\begin{align}
    \ket{\psi_G}=&\left(\prod_{\substack{(j,k)\in\,\\ \text{edges}(G)}}U^\text{(gr)}_{(j,k)}\right)\left(\frac{\ket{0}+\ket{1}}{\sqrt{2}}\right)^{\otimes N},
    \\
& U^\text{(gr)}_{(a,b)}
    =\exp\left(\mathrm{i}\pi |00\rangle\langle 00|_{a,b} \right),
\end{align}
in which case 
$$ U_{\psi}=\left(\prod_{\substack{(j,k)\in\,\\ \text{edges}(G)}}U^\text{(gr)}_{(j,k)}\right) \left(\prod_{\substack{j\in\text{qubits}}}\exp\left(\mathrm{i} \frac{\pi}{4}\sigma^y_j\right)\right).$$ 
Since two-qubit rotations $U^{\text(gr)}_{(j,k)}$ all mutually commute, the coupling $V^{(U_{\psi})}_s(i)$ acts only on spin $i$ and on the spins $j$ whose vertices share an edge with $i$ in the graph $G$. Therefore, this coupling is $(k+1)$-local if there are $k$ edges going out of vertex $i$. Moreover, $V^{(U_{\psi})}_s(i)$ is also geometrically local, if the graph $G$ only connects the qubits which are in geometric proximity. We conclude that for the graphs satisfying the above conditions, a realistic preparation of graph states with local frustration-free steering is possible. Such a protocol can be sped up in the same way it was possible for the product states -- using active feedback via the partial-termination strategy.

For the perfectly mixed starting state, the partial-termination policy gives an \textit{optimal} speed-up of a protocol driven by frustration-free couplings $V^{(U_{\psi})}_s(i)$. Indeed, the protocols in question are then equivalent to an independent set of $N$ $1$-qubit steering protocols (under the unitary transformation $U_{\psi}$). This picture, however, breaks down for a more general starting state. Let us first consider the trivial target $U_{\psi}=\mathbb{I},~\ket{\psi_{\mathrm{targ}}}=\ket{00..0}$, while the starting state is itself entangled (e.g. $\frac{1}{\sqrt{2}}(\ket{00..0}+\ket{11..1})$). In this case, the click received from a single coupling $V_s(i)$ may imply that multiple couplings can be dropped from the applied sequence, and not just $V_s(i)$ itself. This would be more optimal than the partial termination strategy outlined above. The same picture extends to the more interesting case when the target state $\ket{\psi_{\mathrm{targ}}}$ is entangled itself, e.g. a graph state, while the starting state is a product state. Indeed, under the unitary mapping $U_{\psi}$ which takes an entangled state $\ket{\psi_{\mathrm{targ}}}$ to $\ket{00..0}$, the product starting state in turn becomes entangled. Hence, the previous reasoning applies and partial-termination would generally not be an optimal active policy in this situation (although a nontrivial speedup factor $f>1$ is still guaranteed). To accelerate it further, one may apply one of the frustrated-coupling strategies outlined in the following sections.

\subsection{Frustrated system-detector couplings
\label{Sec:IIIB}}

Assuming locality of $\{V_s(\mathbf{p})\}$, for target states other than the product states and states prepared by a shallow circuit, we would generally need to consider non-commuting couplings. For such target states, the first question to tackle is how to design a family of local couplings $\{V_s(\mathbf{p})\}$ that are suitable for a passive protocol. In principle, this can be addressed on a case-by-case basis, tailoring a coupling set to a specific target state. (This approach will be used for the $W$-state preparation in Sec.~\ref{sec:QSM_W_preparation}.) However, this is not always a straightforward task. Therefore, it is interesting to know whether one can devise a general scheme to this end. For this, we propose an approach based on the parent-Hamiltonian construction.

The parent Hamiltonian $H_{\mathrm{par}}(\psi_{\mathrm{targ}})$ is built to have $\ket{\psi_{\mathrm{targ}}}$ as its ground state (we will only consider the non-degenerate case). Assuming that $\ket{\psi_{\mathrm{targ}}}$ hosts a limited amount of entanglement \cite{fannes1992finitely}, $H_{\mathrm{par}}(\psi_{\mathrm{targ}})$ obeys the local projective form \cite{hast06}:
\begin{equation}
\label{eq:parent_Ham}
H_{\mathrm{par}}(\psi_{\mathrm{targ}})=\sum_j H^{(j)}_{\psi}.
\end{equation}
Local projective form means that all terms $H^{(j)}_{\psi}$ are local in real space and that $H^{(j)}_{\psi}$ have $\ket{\psi_{\mathrm{targ}}}$ as their common ground state. The latter property holds even though $H^{(j)}_{\psi}$ in general don't commute with each other --- this is possible due to the ground state degeneracy of $H^{(j)}_{\psi}$. In fact, the ground spaces of $H^{(j)}_{\psi}$ will be central to constructing our coupling family $\{V_s(\mathbf{p})\}$. For the term $H^{(j)}_{\psi}$ nontrivially acting on a collection of $m$ spins, denote its $m$-spin ground states as $\ket{\phi^{(j)}_a}$ and the excited states $\ket{\theta^{(j)}_a}$. 
Given these, we can construct the coupling operators of the following form:
\begin{align}
V^{j}_s(\mathbf{w},&\mathbf{v}, \mathbf{u})=\sum_{ab} w_{ab} \ket{\phi^{(j)}_a}\bra{\theta^{(j)}_b}
\nonumber\\
&+\sum_{ab} v_{ab} \ket{\theta^{(j)}_a}\bra{\theta^{(j)}_b}+
\sum_{ab} u_{ab} \ket{\phi^{(j)}_a}\bra{\phi^{(j)}_b}.
\label{eq:parent_V}
\end{align}
A particular example of this construction will be addressed in detail for the Matrix Product State target states in Sec.~\ref{sec:quantum_compass} (see also Refs.~\cite{kraus2008preparation, roy2020measurement}). 

For a generic (fixed) value of parameters $(\mathbf{w},\mathbf{v}, \mathbf{u})$, a passive protocol driven by the couplings $V^{j}_s(\mathbf{w},\mathbf{v}, \mathbf{u})$ does converge to $\ket{\psi_{\mathrm{targ}}}$. In particular, alternating different $j$ in $V^{j}_s(\mathbf{w},\mathbf{v}, \mathbf{u})$ at consecutive protocol cycles allows to steer the system to $\ket{\psi_{\mathrm{targ}}}$ as the joint ground space of all couplings $H^{(j)}_{\psi}$ (see \cite{kraus2008preparation} for a related statement proven rigorously for AKLT model). As long as the conditions on $\ket{\psi_{\rm{targ}}}$ for locality of $H_{\mathrm{par}}(\psi_{\mathrm{targ}})$ are satisfied, this concludes the construction of an appropriate coupling set for $\ket{\psi_{\rm{targ}}}$. 

Let us now consider an active-protocol construction. Unlike in the frustration-free protocol, the operators $V^{j}_s(\mathbf{w},\mathbf{v},\mathbf{u})$ for different values of $j$ generally don't commute. Therefore, the measurement outcome of steering by $V^{j}_s(\mathbf{w},\mathbf{v}, \mathbf{u})$ impacts the outcomes of steering at locations close to $j$. As a result, the partial-termination strategy cannot be applied to this coupling set, as it assumes that the respective cycles of the protocol can be considered separately. Due to this difficulty, we classify steering with noncommuting couplings as frustrated \cite{footnote1}. The feedback strategy for frustrated steering should continuously coordinate the application of different couplings in the protocol. In a many-body context, this becomes a complicated navigation-type problem (cf. Ref. \cite{lar20}). We devote the following two sections to the study of such possible Hilbert-space navigation policies.

\section{Quantum compass:\\ Cost-function policies
\label{sec:quantum_compass}}

One way to enable the Hilbert-space navigation is to introduce a cost function $C(\rho)$, which is to be minimized in the protocol. The basic example would be the infidelity $C(\rho)=R(\rho,\ket{\psi_{\mathrm{targ}}})$ of the system state $\rho$ to the target state $\ket{\psi_{\mathrm{targ}}}$, defined in Eq.~(\ref{eq:infidelity}).
Achieving the global minimum $R(\rho,\ket{\psi_{\mathrm{targ}}})=0$ of this cost function would be equivalent to preparing the target state. 
In general, to calculate $R(\rho,\ket{\psi_{\mathrm{targ}}})$, one needs to know the state of the system $\rho$. This is, in principle, feasible, as we control the system evolution given all measurement outcomes and therefore can numerically simulate it in parallel to the experiment.
However, the requirement of such a simulation being done in parallel to the experiment puts a restriction on the size of the system that one can work with. For now, we will accept this limitation; finding ways to mitigate it is among the worthwhile potential extensions of this work.

With a given cost function $C(\rho)$ at hand, we can use it to form the active decision for the coupling operator $V_s(\mathbf{p})$. 
The ultimate strategy is to pick $V_s(\mathbf{p})$ which brings the system to the global minimum $\ket{\psi_{\mathrm{targ}}}$ in the fastest expected time.
For $C(\rho)=R(\rho,\ket{\psi_{\mathrm{targ}}})$ this is equivalent to the ultimate strategy defined by dynamic programming \cite{fu2014feedback}, requiring unrealistic computation power.
Instead, one can use a cheaper approach to cost function minimization -- the ``greedy strategy.'' Specifically, one can use $V_s(\mathbf{p})$ that yields the fastest expected reduction of the cost function in a single step 
of the evolution:
\begin{equation}
    V^\text{(greed)}_s(\mathbf{p})
    =\mathrm{argmin}_{V_s(\mathbf{p})}
    R[\Lambda_{V_s(\mathbf{p})}(\rho)],
\end{equation}
where $\Lambda_{V_s(\mathbf{p})}(\rho)$ is defined in Eq.~\eqref{eq:steering_map}.
If there are multiple minima, we will postulate that $\mathrm{argmin}$ returns a random representative among those. With only a small amount of computations needed to decide for the optimal next coupling $V^\text{(greed)}_s(\mathbf{p})$, this greedy procedure allows us to avoid the complex long-term analysis of the protocol.

\subsection{Greedy steering: Matrix Product States}
\label{sec:MPS_preparation}

As one can see numerically from a direct implementation, the greedy minimization of the cost function can accelerate the state preparation by a large factor. To demonstrate this, we consider the Matrix Product State \cite{perez06} (MPS) targets.
We focus on spin-1 uniform MPS with bond dimension 2, which are defined as follows:
\begin{equation}
    \ket{\psi_{\mathrm{MPS}}(A)}=\sum^3_{\alpha_k=1} \sum^{2}_{i_k=1} A^{\alpha_1}_{i_1i_2}A^{\alpha_2}_{i_2i_3}..A^{\alpha_N}_{i_N i_1} \ket{\alpha_1\alpha_2..\alpha_N},
    \label{eq:MPS_def}
\end{equation}
where $A$ in each factor is the same tensor. One example of MPS $\ket{\psi_{\mathrm{MPS}}(A)}$ that we consider is the Affleck-Lieb-Kennedy-Tasaki (AKLT) state, with:
\begin{align}
A^1_{ij}=&{\scriptscriptstyle \begin{pmatrix}0 & \sqrt{\frac{2}{3}}\\ 0 & 0 \end{pmatrix}_{ij}},~~ A^2_{ij}={\scriptscriptstyle\begin{pmatrix}-\sqrt{\frac{1}{3}} & 0 \\ 0 & \sqrt{\frac{1}{3}} \end{pmatrix}_{ij}},\\ &A^3_{ij}={\scriptscriptstyle\begin{pmatrix}0 & 0 \\ - \sqrt{\frac{2}{3}} & 0 \end{pmatrix}_{ij}}.
\end{align}
To test the generality of our approach, we also consider random Matrix Product States as targets. In particular, the tensor $A$ is to be generated at random, from a uniform distribution $A^{\alpha_k}_{i_ki_{k+1}}\sim {\cal U}_{[0, 1]}$ (subsequently normalizing the resulting state $\ket{\psi_{\mathrm{MPS}}(A)}$).

To design a measurement-driven preparation protocol for $\ket{\psi_{\mathrm{MPS}}(A)}$, note that a generic MPS of the form Eq.~\eqref{eq:MPS_def} admits a parent Hamiltonian with individual terms acting on pairs of sites:
\begin{align}
    H_\mathrm{par}(A)&=\sum_{k} h^{k,k+1},~~N+1\equiv1,\\h^{k,k+1}&=\mathbb{I}-\sum_{i_k i_{k+2}}\mathbb{P}(\sum_{i_{k+1}}A^{\alpha_k}_{i_k i_{k+1}}A^{\alpha_{k+1}}_{i_{k+1} i_{k+2}}),
\end{align}
where $\mathbb{P}(v^{\alpha_k,\alpha_{k+1}})$ is the projector onto the state $\sum_{\alpha_k,\alpha_{k+1}} v^{\alpha_k,\alpha_{k+1}}\ket{\alpha_k \alpha_{k+1}}$. Defined as such, each term $h^{k,k+1}$ has $\ket{\psi_\mathrm{MPS}(A)}$ as its ground state. For the AKLT state, $H_\mathrm{par}(A)$ coincides with the AKLT Hamiltonian \cite{affleck1987rigorous}.

As discussed in Sec.~\ref{Sec:IIIB}, the parent Hamiltonian $H_\mathrm{par}(A)$ admits 
a set of couplings $V(\mathbf{p})$ defined from the local spectrum of $h^{k,k+1}$. Denoting the $4$ ground states and $5$ excited states of $h^{k,k+1}$ as $\ket{\phi^{(j)}_a}$ and $\ket{\theta^{(k)}_b}$ respectively, these couplings can be defined as in Eq.~\eqref{eq:parent_V} -- we will denote them as $V^{k,k+1}_s(\mathbf{w},\mathbf{v}, \mathbf{u})$. 
In a passive protocol, we will cyclically alternate between different sites $k=1,..N$, while drawing $\mathbf{w},\mathbf{v}$, and $\mathbf{u}$ at random for each link (each matrix element being a Gaussian random variable with zero mean and unit variance). We demonstrate numerically that this passive protocol guarantees preparation of the target state in a finite number of time-steps \cite{footnote2}. For an active feedback strategy to be used on top of this, we propose a greedy policy relative to $C(\rho)=R(\rho,\ket{\psi_{\mathrm{targ}}})$ to select $\mathbf{w},\mathbf{v}$, and $\mathbf{u}$.  The key question is -- does such an active policy yield any speedup relative to the passive protocol? 

\begin{figure}[H]
\begin{center}
\includegraphics[width=0.8\linewidth]{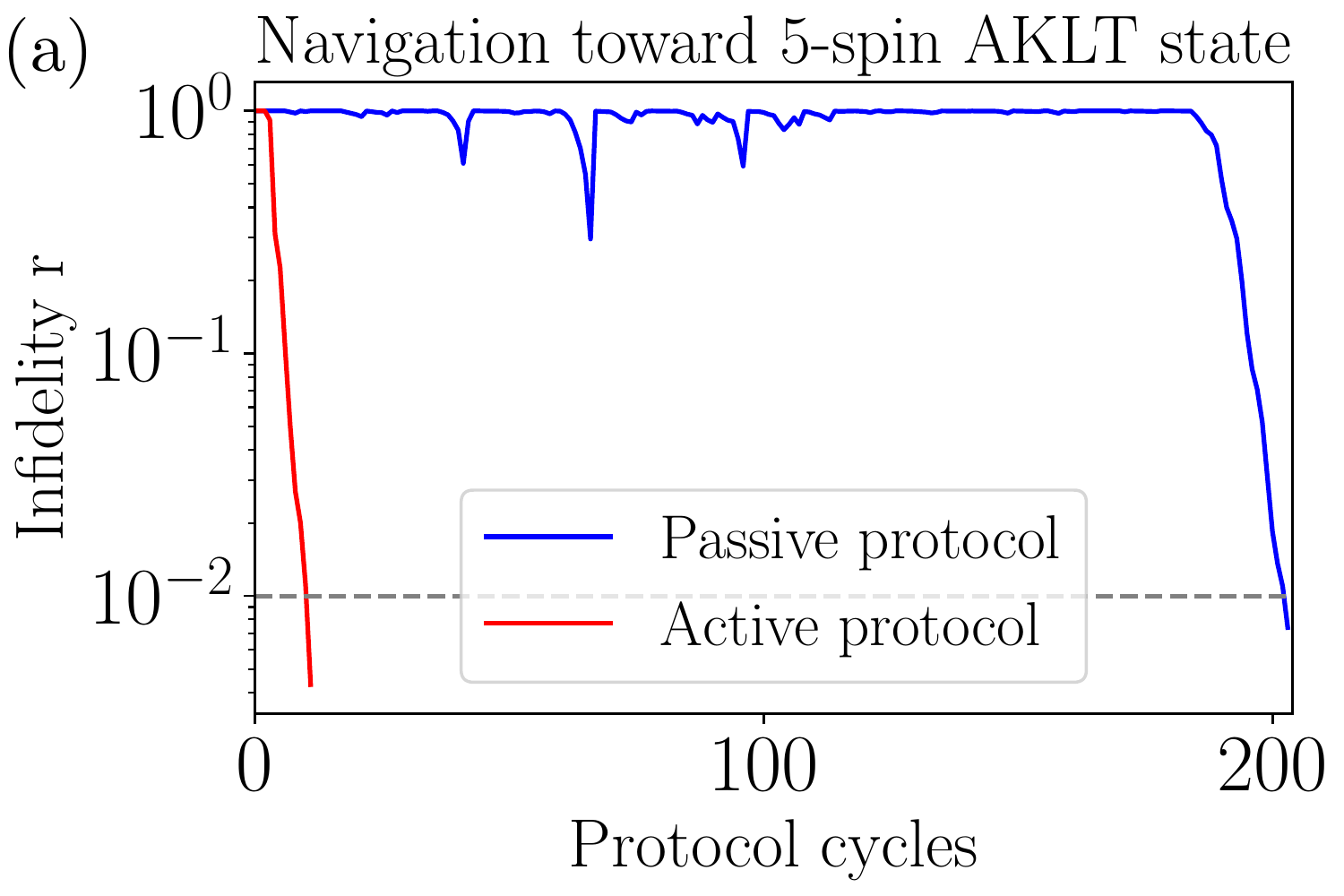}
\vspace*{-0.25cm}
\includegraphics[width=0.8\linewidth]{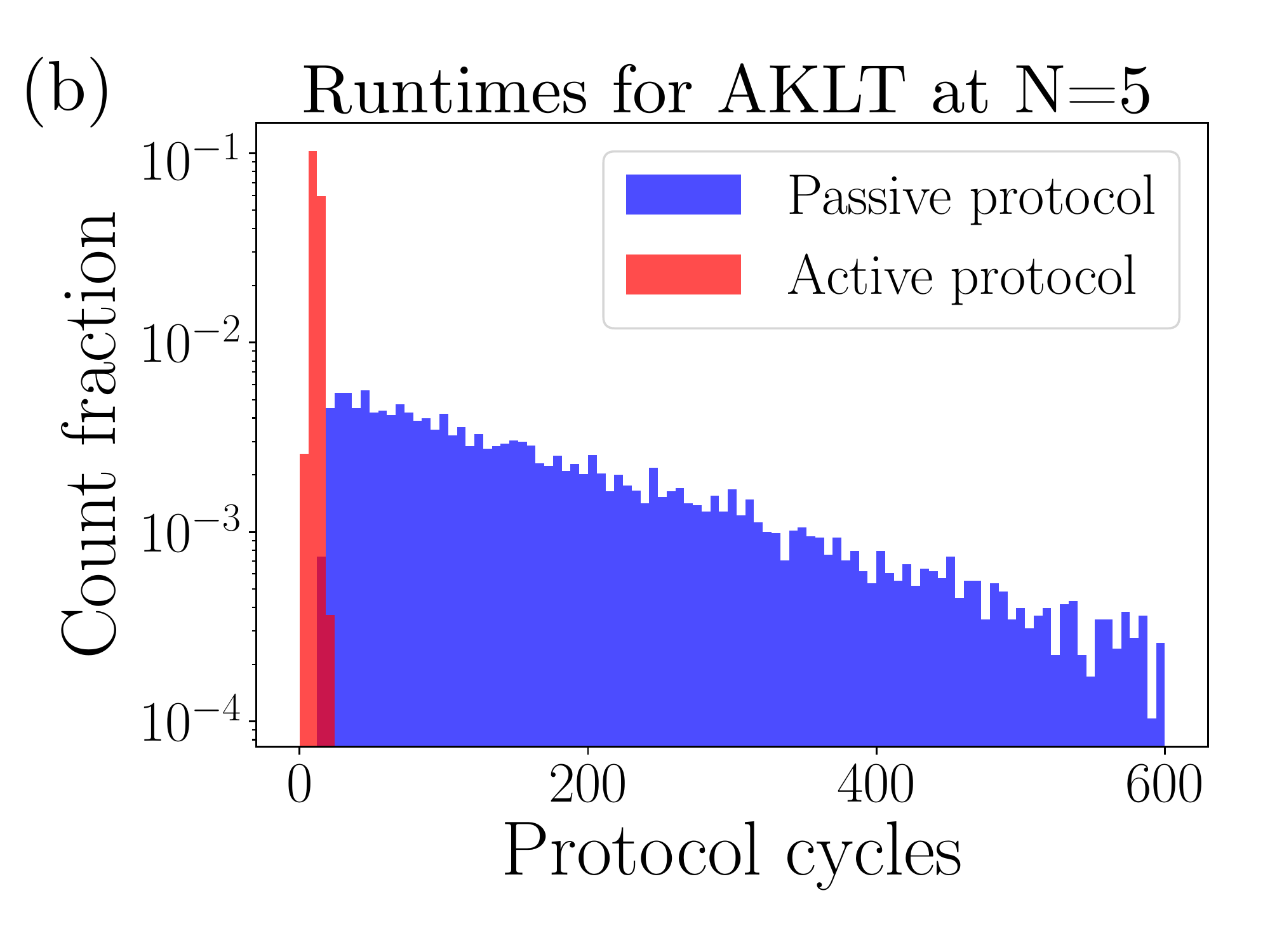}
\vspace*{-0.25cm}
\includegraphics[width=0.8\linewidth]{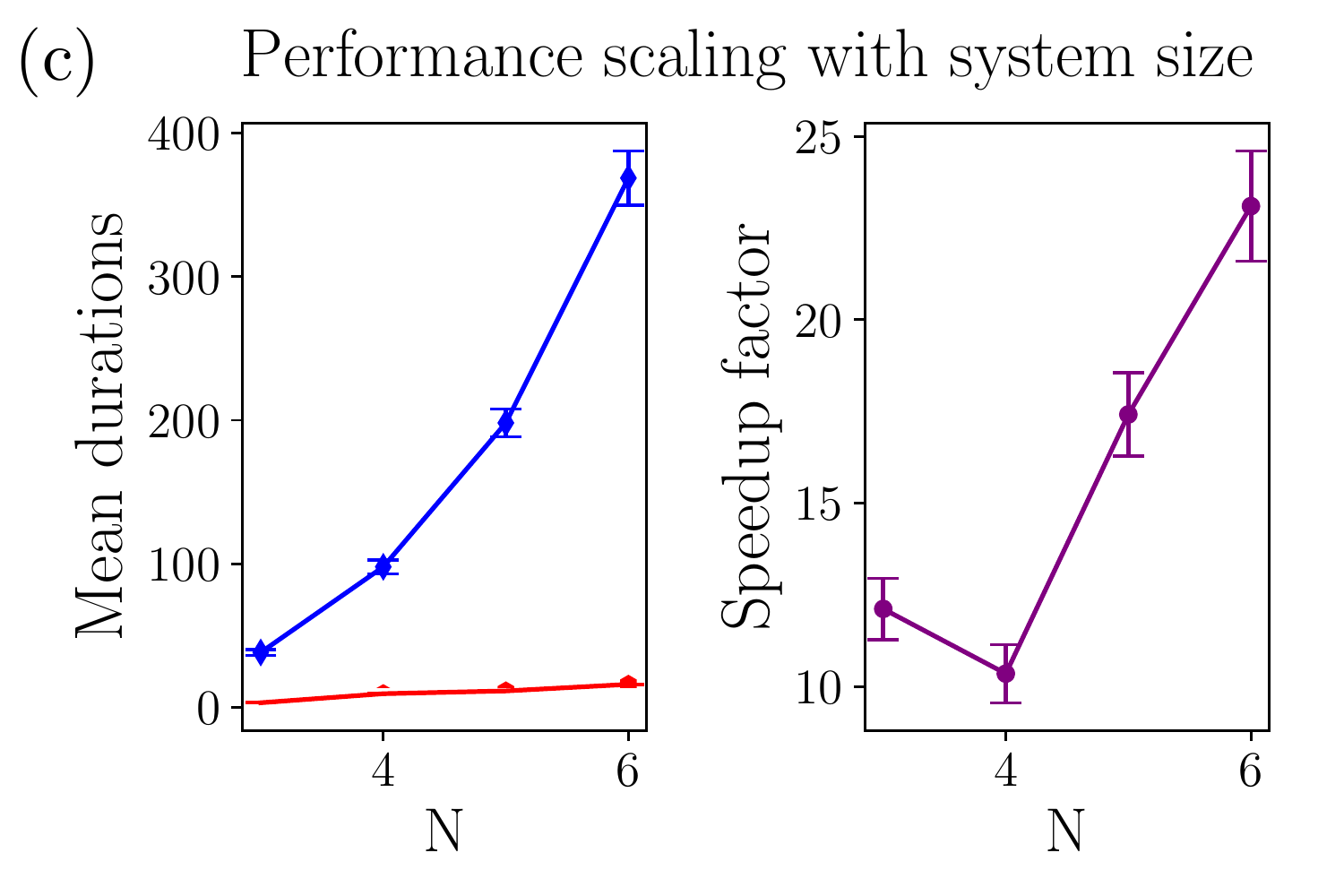}

\end{center}
\vspace*{-0.5cm}
\caption{\label{fig:AKLT} 
Simulated measurement-driven preparation of the AKLT state up to infidelity $R<\epsilon = 0.01$.
(a) Infidelity as a function of the protocol cycle for active and passive protocol runs towards a 5-spin AKLT state. These example runs are characterized by the runtimes similar to the average runtimes of respective protocols ($\tau^\mathrm{(pas)}_\mathrm{av}\simeq 200$, $\tau^\mathrm{(act)}_\mathrm{av}\simeq 11$). The passive protocol experiences setbacks in its performance at a few moments, with infidelity first decreasing and then resetting back to $1$. The active protocol manages to avoid this issue. (b) Histograms of protocol runtimes $\tau$ for the five-spin AKLT state preparation. An exponential decaying profile, characteristic of a Poissonian process, can be clearly observed for the passive protocol (note the log scale). All recorded runs for an active protocol lasted far less than average passive runtime $\tau^\mathrm{(pas)}_\mathrm{av}\simeq 200$. Both histograms were compiled from $10^4$ simulated runs; the figure is truncated at $600$ cycles for better presentation. (c) Scaling of the active protocol's advantage with system size $N$. The speedup factor $f$ tends to increase significantly as the system scales, with factor $23$ being the estimated $f$ at $6$ spins. The error bars represent $95\%$  
confidence intervals due to sampling error in numerical simulation. A sample of $10^3$
runs was collected to simulate the performance of both active and passive protocols at each system size.}
\end{figure}

\begin{figure}[H]
\begin{center}
\includegraphics[width=\linewidth]{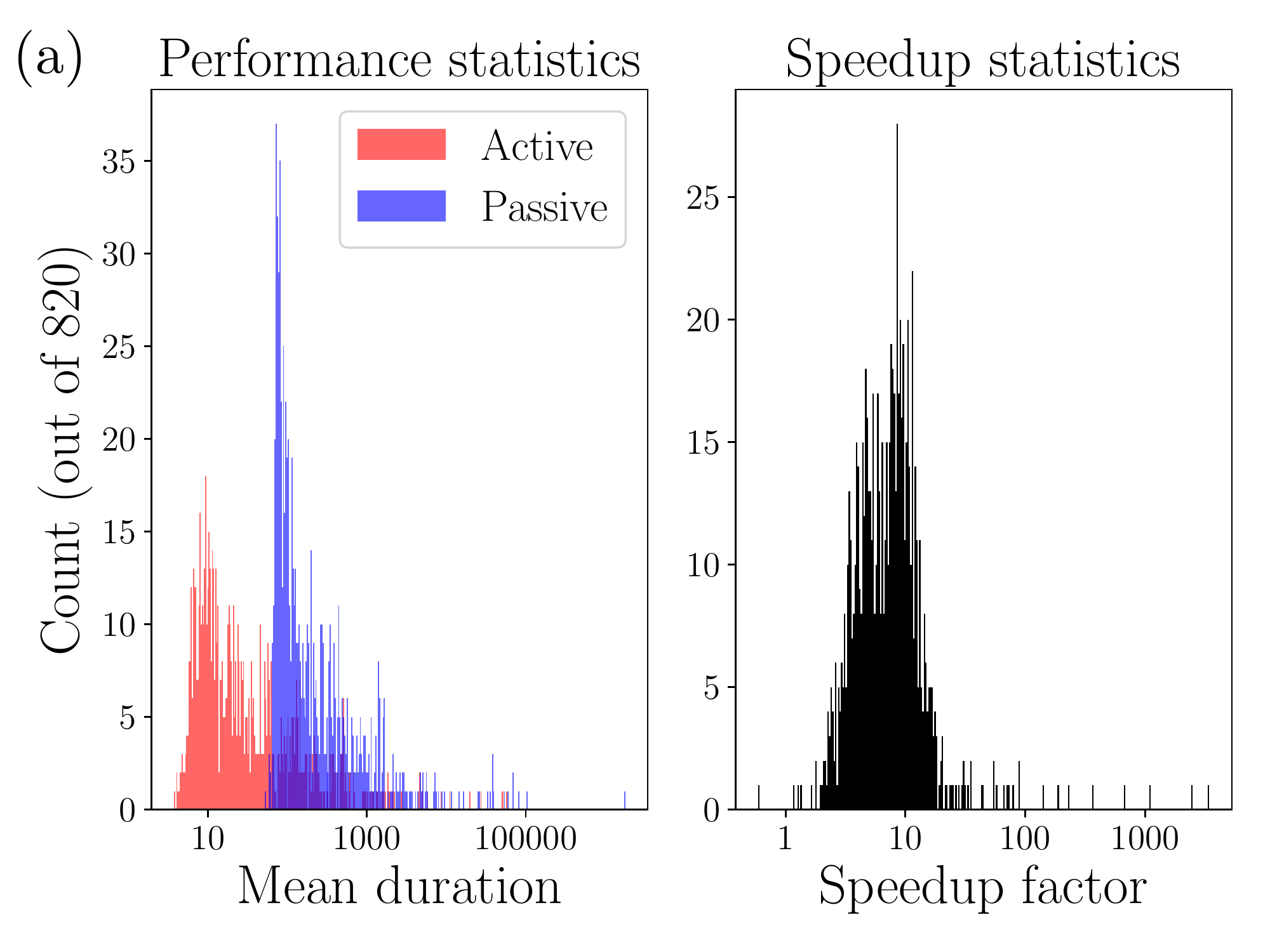}
\includegraphics[width=\linewidth]{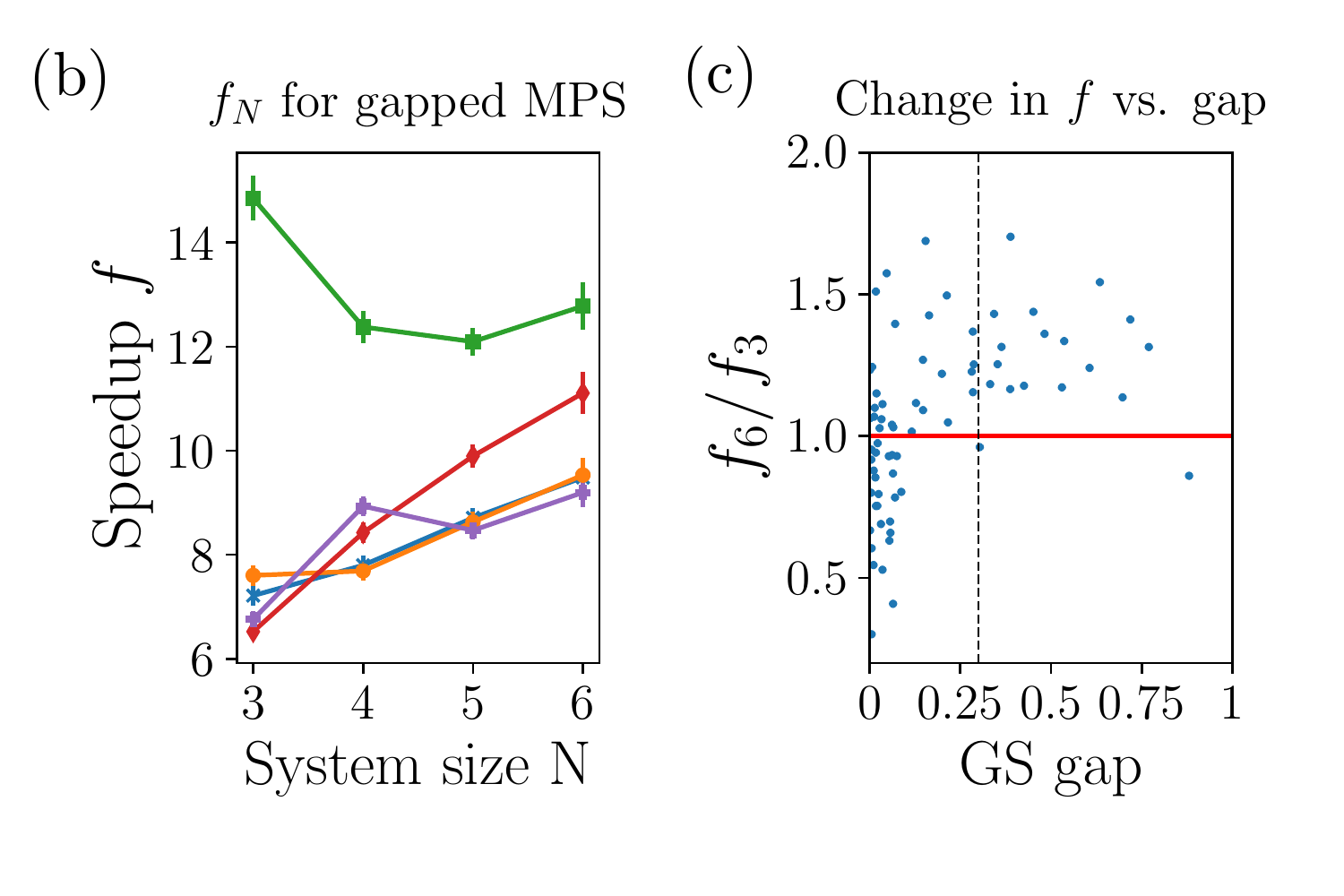}

\end{center}
\vspace*{-0.25cm}
\caption{\label{fig:rMPS} 
Simulations of measurement-based preparation of random Matrix Product State with error $R<\epsilon = 0.01$.
(a) Histograms of the average performance metrics for the preparation of the five-spin random Matrix Product State. Displayed are $\tau^\mathrm{(pas)}_\mathrm{av}$, $\tau^\mathrm{(act)}_\mathrm{av}$, and the resulting speedup $f=\tau^\mathrm{(pas)}_\mathrm{av}/\tau^\mathrm{(act)}_\mathrm{av}$. The results depend strongly per random instance of $A$ (note the log scale). Among the sampled instances, $f$ is greater than $1$ for all instances but one; on the high end, $f$ ranges up to $f\simeq3400$. Most counts are around $f\sim 10^1$, and the average value is $f\simeq19$. Each histogram was compiled from $10^3$ simulated runs for $820$ randomly generated random Matrix Product States (cf. main text). (b) Scaling of the active protocol's advantage with system size $N$ for target MPS with a large parent Hamiltonian gap. The target states are randomly generated while postselecting for the gap to be larger than a practical threshold value of $0.3$ at $N=5$. For most such targets, the speedup factor tends to increase significantly as the system scales. (cf. Fig.~\ref{fig:AKLT}c for AKLT model, whose spectral gap at $N=5$ is equal to $0.45$.) Each data point is obtained via sampling over $10^4$ runs of the active and passive versions of the same protocol. Error bars display the respective $95\%$ confidence interval. 
(c) Correlation between the speedup factor at $N=3$ and $N=6$ for random MPS targets with the respective parent Hamiltonian gaps. The vertical dotted line is put at the threshold gap value of $0.3$. For most targets with a larger gap, the speedup $f$ increases with system size. Each gap value is calculated at $N=5$.} 
\end{figure}

Our numerics demonstrates that MPS targets typically admit a significant value of speedup factor $f=\tau^\mathrm{(pas)}_\mathrm{av}/\tau^\mathrm{(act)}_\mathrm{av}$.
For a particular case of AKLT target (Fig.~\ref{fig:AKLT}), the speedup factor ranges between $f\simeq 10$ and $f\simeq 23$ for system sizes $3$ to $6$.
For random MPS as target states, $f$ strongly fluctuates per $A$; to get an idea of the statistics of $f$ across the targets, we perform an extensive numerical study (Fig.~\ref{fig:rMPS}). Among $820$ random MPS targets sampled at $N=5$ (Fig.~\ref{fig:rMPS}a), the speedup ranges up to $f_{\mathrm{max}}\simeq 3400$, with the average value $f_{\mathrm{av}}\simeq 19$. In other places of the manuscript we refer to these as $f_{\mathrm{max}}\sim 10^3$ and $f_{\mathrm{av}}\sim 10^1 $, to focus on the orders of magnitude and not the specific values. Such a significant speedup across many MPS targets underscores the potential of applying the cost function-based active policy to a wide variety of settings. 

\pagebreak

An additional point of interest is the change of $f$ with the system size (Fig.~\ref{fig:rMPS}b,c). Numerics demonstrates that $f$ may either tend to increase or decrease with the system size $N$, depending on the tensor $A$ of the target MPS. In particular, the direction of this tendency is highly correlated with the ground state gap of the respective parent Hamiltonian, as demonstrated in Fig.~\ref{fig:rMPS}c for randomly generated $A$ tensors. For a large enough gap, the clear majority of targets displays an increase in speedup with system size. This is further illustrated Fig.~\ref{fig:rMPS}b, which focuses on $f$ as a function of $N$ for targets with a large gap in respective parent Hamiltonians. The growth of $f$  with $N$ for such a broad family of targets is another compelling feature of the active policy presented here.

\subsection{Landscape exploration and alternative cost functions}
\label{sec:orthogonality_catastrophe}

The greedy approach defined above does not suffer from the presence of local minima in the cost-function landscape, which are typical obstacles in optimization procedures. In our approach, the target state is a dark state for each steering operator. Therefore the infidelity of the running state to the target state never increases on average. This fact and the convergence of the respective passive protocol guarantee the convergence of the greedy optimization protocol. On the other hand, there is no formal guarantee that our active protocol yields a speedup compared to the passive one. Instead, we note that the speedup factor $f>1$ is achieved \textit{in practice} for most examples that we tested. This includes the AKLT state targets, as well as 819 out of 820 random MPS targets that we considered (Fig.~\ref{fig:rMPS}a). Obtaining formal guarantees may also be possible under some conditions on a target state; this could be an interesting direction for future investigation.

Despite the absence of ``glassy'' landscape, our greedy strategy still harbors a potential challenge.
For the greedy procedure to be effective, it should always yield a nonzero bias in favor of a specific
$V^\text{(greed)}_s(\mathbf{p})$ (or a small subset thereof). In other words, the landscape of the given cost function $C(\rho)$ should not be flat. It follows that applying the infidelity measure $R(\rho,\ket{\psi_{\mathrm{targ}}})$ is in general a flawed strategy. The reason is that a $(2^N-1)$-dimensional subspace of states in the $N$-body Hilbert space is orthogonal to the target state. Consider the case when the starting state belongs to that subspace. The state would in general stay in this subspace after a single steering cycle with a local coupling $V_s(\mathbf{p})$. For our purposes, it implies that the infidelity measure $R$ is equal to $1$ for a large manifold of states, and there might be no direction of increase that would allow us to choose an appropriate coupling. 
In this scenario, the active steering is effectively reduced to a passive one (with no preference with respect to choosing a particular coupling).
Since we assume the convergence of the passive protocol, such (``local'') flatness scenarios do not disable the convergence of the active protocol. However, this might still diminish the resulting speedup factor $f$. 

The most direct example of this effect of flat landscapes can be observed when applying the greedy policy to frustration-free steering
(see Sec.~\ref{sec:mutually_commuting}). 
For simplicity, let us again take the product state of $N$ qubits $\ket{00..0}$ as the target state, the state $\ket{11..1}$ as the starting state, and the couplings $V(i)=\sigma^-_i$ for steering. Only after such steering protocol results in $N$ successful click events, $R(\rho,\ket{\psi_{\mathrm{targ}}})$ gains a nonzero value. Thus before $N-1$ clicks, the greedy policy for $R^\text{(inf)}$ will not be capable of providing a biased decision for the next coupling. Strongly enhanced by the system size, this phenomenon is reminiscent of Anderson's orthogonality catastrophe \cite{anderson1967infrared}. 

As a remedy to this deficiency, the full target-state fidelity can be replaced with its more localized versions. For example, a ``subsystem infidelity'' measure can be introduced:
\begin{equation}
    R_{\mathcal{S}}(\rho,\ket{\psi_{\mathrm{targ}}})=\sum_{\sigma\in \mathcal{S}} \left[1-
    \mathrm{tr}\left(\sqrt{\sqrt{\rho_{\mathrm{targ},\sigma}} \rho_{\sigma} \sqrt{\rho_{\mathrm{targ},\sigma}} }\right)^2 \right],
\end{equation}
where $\rho_{\mathrm{targ},\sigma}$ ($\rho_{\sigma}$) is the reduced density matrix of the target state (current state) with respect to subsystem $\sigma$. $\mathcal{S}$ is the family of subsystems from which $\sigma$ are drawn; the choice of $\mathcal{S}$ depends per target state.
In the case of the $\ket{11..1}\rightarrow\ket{00..0}$ protocol described above, the appropriate $\mathcal{S}$ would be the set of individual spins. Unlike $R$, such quantity $R_{\mathcal{S}}$ changes every time when a click occurs in this protocol. As a result, the greedy policy with respect to the local $R_{\mathcal{S}}$ would yield the partial-termination protocol of Sec.~\ref{sec:mutually_commuting}, significantly boosting the preparation of such a product state.  

By continuity with the case of the product state target, such preference for $R_{\mathcal{S}}$ should extend to the weakly-entangled target states, and maybe to some highly-entangled targets. However, we did not see a manifestation of this in our MPS simulations, where using $R_{\mathcal{S}}$ as a cost function did not yield any improvement compared to $R$. As a likely explanation for this, the orthogonality catastrophe should become manifest only at large system sizes, where the classical simulation of the protocol is also hindered. However, we expect that some practical target states may still develop a noticeable performance difference between $R_{\mathcal{S}}$ and $R$, similarly to the case of the product state target. A further study of this question constitutes a promising direction for future work.

In addition to modifying the cost function, another way to fix the landscape flatness issue is to move away from the cost-function based policies entirely. This is one of the key motivations for an alternative (Quantum State Machine) framework we introduce in the next Section. Such an alternative framework can indeed outperform the cost function-based policy due to the landscape flatness issue the latter occasionally experiences. An example of this is shown in Sec.~\ref{sec:QSM_W_preparation}, where the measurement-based preparation of the W-state is investigated.

\subsection{Role of measurement imperfections}
\label{failed-active}

The above greedy policy is formulated for an ideal case of perfect detectors. Reducing ``detection efficiency'' (say, recording a click readout instead of an actual no-click one) in the active protocol could eventually reduce it to a passive one. Indeed, the choice of the further couplings will be based on a wrong position in the cost-function landscape, and, hence, might become completely random with respect to the actual landscape. Nevertheless, in our work, the set of couplings still guarantees successful passive steering. Therefore, the speedup factor for the active protocol remains generically higher than 1, even in the presence of such errors. 

One can roughly describe the crossover in the speedup factor between the ideal and ``imperfect'' active protocols by introducing a probability of a ``failed measurement'' and assuming a fully passive steering after that particular measurement step. 
Given a typical value of the error (``measurement imperfection'') time $\tau^\text{(err)}$ for switching from active to passive protocol, one can estimate the ``failed'' speedup factor as follows. For simplicity, we assume that the speedup factor in the active protocol is time-independent (for sufficiently long times), i.e., the distance to the target state for a given time $\tau$ in the active protocol is the same as the distance for time $f\,\tau$ in the passive one. This means that, for an arbitrary initial state and the target infidelity, the typical active-steering time, $\tau^\text{(act)}$, and passive-steering time, $\tau^\text{(pas)}$, are related by the same ``ideal'' speedup factor $f$. This assumption is in a good agreement with our numerical results, where the speedup factor is roughly independent of the target precision.
Let us further specify that, for the passive protocol, the distance to the target state depends on time exponentially (this is the case when passive steering is described by a gapped Lindbladian, as, e.g., in the case of the AKLT model \cite{roy2020measurement}).  Then, the total time to reach the target state for such corrupted active steering is given by $\tilde{\tau}^\text{(act)}=\tau^\text{(pas)}-(f-1) \tau^\text{(err)}$ (with $\tau^\text{(err)}<\tau^\text{(act)}$, otherwise active steering is essentially unaffected by errors). The effective speedup factor thus becomes $$\tilde{f}=\frac{\tau^\text{(pas)}}{\tilde{\tau}^\text{(act)}}
=\frac{f \tau^\text{(act)}}{f \tau^\text{(act)}-(f-1)\tau^\text{(err)}}>1.$$ 
Note that this speedup factor now depends on the target precision through $\tau^\text{(act)}$. 

One can, in principle, refine this estimate, by using the full distribution function of the runtimes and error times, or by relaxing the assumption of time-inhomogeneity of the speedup factor. Further, it is interesting to study a host of non-fatal errors, when the location of the system in Hilbert space is only slightly blurred. In addition, it is noteworthy that possible imperfections in active protocols can be monitored given full information on the readouts, and the corresponding error-correction strategies can be designed. We relegate this and related questions to future work.

\section{Hilbert-space orienteering map: Quantum State Machine
\label{sec:QSM}}

In this section, we present an orienteering tool that is an alternative to cost-function minimization: mapping out the steering transformations with a Quantum State Machine (QSM) construction. We then illustrate navigation in many-body Hilbert space, employing this machinery to the preparation of the highly entangled W-state of three qubits.

\subsection{QSM generalities}
\label{sec:QSM_generalities}

Consider the transformation of the system's state,  $\Lambda^\text{(cl)}_{V_{s}}$ and $\Lambda^\text{(ncl)}_{V_s}$, associated to steering with a specific coupling $V_s$ in a given readout scenario (click or no-click, respectively, see Eqs.~\eqref{eq:click_scenario_map} and \eqref{eq:no-click_scenario_map}). 
Every such steering transformation conserves the purity of the state. Therefore, it is convenient to encode transformations $\Lambda^\text{(cl, ncl)}_{V_s}$ in their action on Hilbert space basis states $\ket{\phi_\alpha}$:
\begin{align}
    & \Lambda^\text{(cl,ncl)}_{V_s}(\ket{\phi_\alpha})=\frac{1}{\sqrt{p^\text{(cl,ncl)}}}\sum_\beta L_{\alpha\beta}^\text{(cl,ncl)} \ket{\phi_\beta} \label{eq:pure_basis_action}\\
    & L_{\alpha\beta}^\text{(cl)}=\bra{\phi_\beta}\delta t V_s\ket{\phi_\alpha}, \label{eq:basis_click_action}\\
    & L_{\alpha\beta}^\text{(ncl)}=\bra{\phi_\beta}1-\delta t^2 V^\dag_s V_s/2
    \ket{\phi_\alpha},
    \label{eq:basis_noclick_action}
\end{align}
where $p^\text{(cl)}$ ($p^\text{(ncl)}$) is the probability of a click (non-click) readout upon this steering action. Note that in Eq.~\eqref{eq:pure_basis_action}, we extended the action of $\Lambda_{V_s}$ to pure states by a slight abuse of notation compared to Eq.~(\ref{eq:click_scenario_map}).

\begin{figure}[t]
\includegraphics[width=0.9\linewidth]{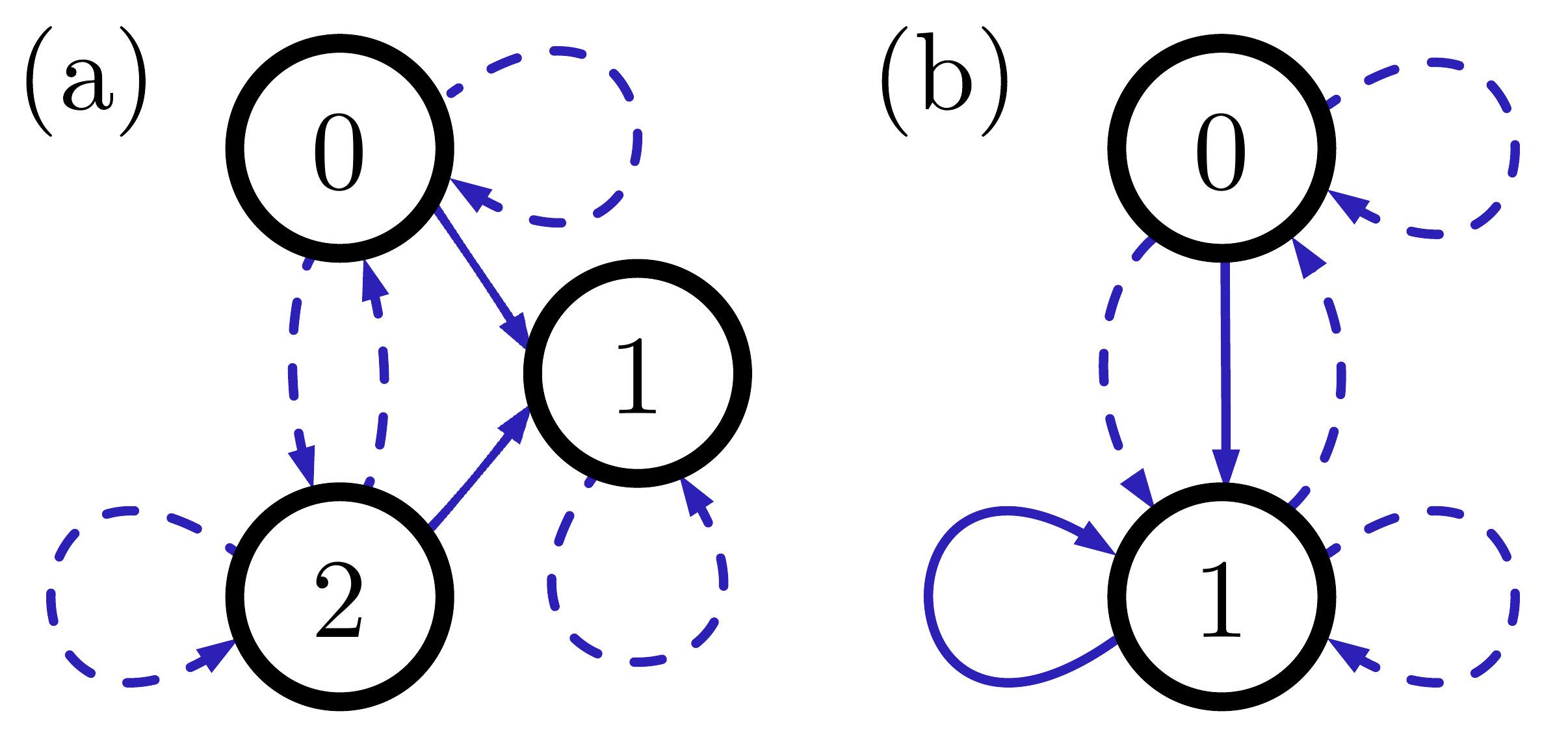}
\caption{\label{fig:QSM_noclick} Examples of steering graphs (see definition in Sec.~\ref{sec:QSM_generalities}): (a) Steering graphs on a 3-level system, corresponding to the coupling $V_s=\gamma(\ket{1}\bra{0}+\ket{1}\bra{2})$. Graph $L^\text{(cl)}$  for click action is depicted with solid arrows and the graph $L^\text{(ncl)}$ for no-click action with dashed arrows. Due to the identity operator in Eq.~\eqref{eq:basis_noclick_action}, every vertex is decorated with a self-loop from the $L^\text{(ncl)}$ graph. To see how the rest of $L^\text{(ncl)}$ can be deduced from $L^\text{(cl)}$, consider the example of $e^{(ncl)}_{02}$ (dashed arrow from state 0 to 2). According to the graphical approach from Sec.~\ref{sec:QSM_generalities}, one is to follow edge $e^{(cl)}_{01}$ (solid arrow from 0 to 1) forward and then $e^{(cl)}_{21}$ (solid arrow from 2 to 1) backward - and thus manages to travel from state 0 to 2, in correspondence to $e^{(ncl)}_{02}$. 
(b) Steering graphs on a 2-level system, as defined by the coupling $V_s=\gamma(\ket{1}\bra{1}+\ket{1}\bra{0})$. Following the same rule as above, inter-vertex edges of $L^\text{(ncl)}$ can be deduced from $L^\text{(cl)}$. For example, by following the edge $e^\text{(cl)}_{11}$ forward and then the edge $e^\text{(cl)}_{01}$ backward, one performs a transition from state 1 to state 0, thus reproducing the edge $e^\text{(ncl)}_{10}$ from $L^\text{(ncl)}$. 
}
\end{figure}

Amplitudes $L_{\alpha\beta}^\text{(cl,ncl)}$ allow to represent the steering action $\Lambda^\text{(cl,ncl)}_{V_s}$ as a graph. The vertices in such a \textit{steering graph} correspond to the Hilbert space basis states, and the edges describe the steering transformations. The edges are directed and weighted with complex amplitudes. Specifically, an edge $\alpha \rightarrow \beta$ is to be weighted with amplitude $L_{\alpha\beta}^\text{(cl, ncl)}$ (edges weighted with zero amplitudes are excluded from the graph). Implying this definition, we will use the notation $L^\text{(cl, ncl)}$ for the steering graphs themselves. For basic examples of steering graphs, please refer to Fig.~\ref{fig:QSM_noclick}. 

Since the weights $L_{\alpha\beta}^\text{(cl)}$ are proportional to the matrix elements of coupling operator $V_s$ while $L_{\alpha\beta}^\text{(ncl)}$ can be expressed via $V_s$ as well, the graph $L^\text{(ncl)}$ for the no-click action can be inferred entirely from the graph $L^\text{(cl)}$ for the click action. In particular, due to the term $\propto V^\dag_s V_s$, graph $L^\text{(ncl)}$ contains an edge $e^\text{(ncl)}_{\alpha\beta}$ from vertex $\alpha$ to $\beta$, if a graph $L^\text{(cl)}$ contains edges $e^\text{(cl)}_{\alpha\mu}$ and $e^\text{(cl)}_{\beta\mu}$ for any vertex $\mu$ (see Fig.~\ref{fig:QSM_noclick}). Heuristically speaking, to yield a $L^\text{(ncl)}$-edge, one has to first follow a $L^\text{(cl)}$-edge forward, and then another $L^\text{(cl)}$-edge backward. Furthermore, due to the additional identity operator term in Eq.~\eqref{eq:basis_noclick_action}, any graph for the no-click steering action will also include self-loops on each vertex. 

\begin{figure}[t]
\includegraphics[width=0.6\linewidth]{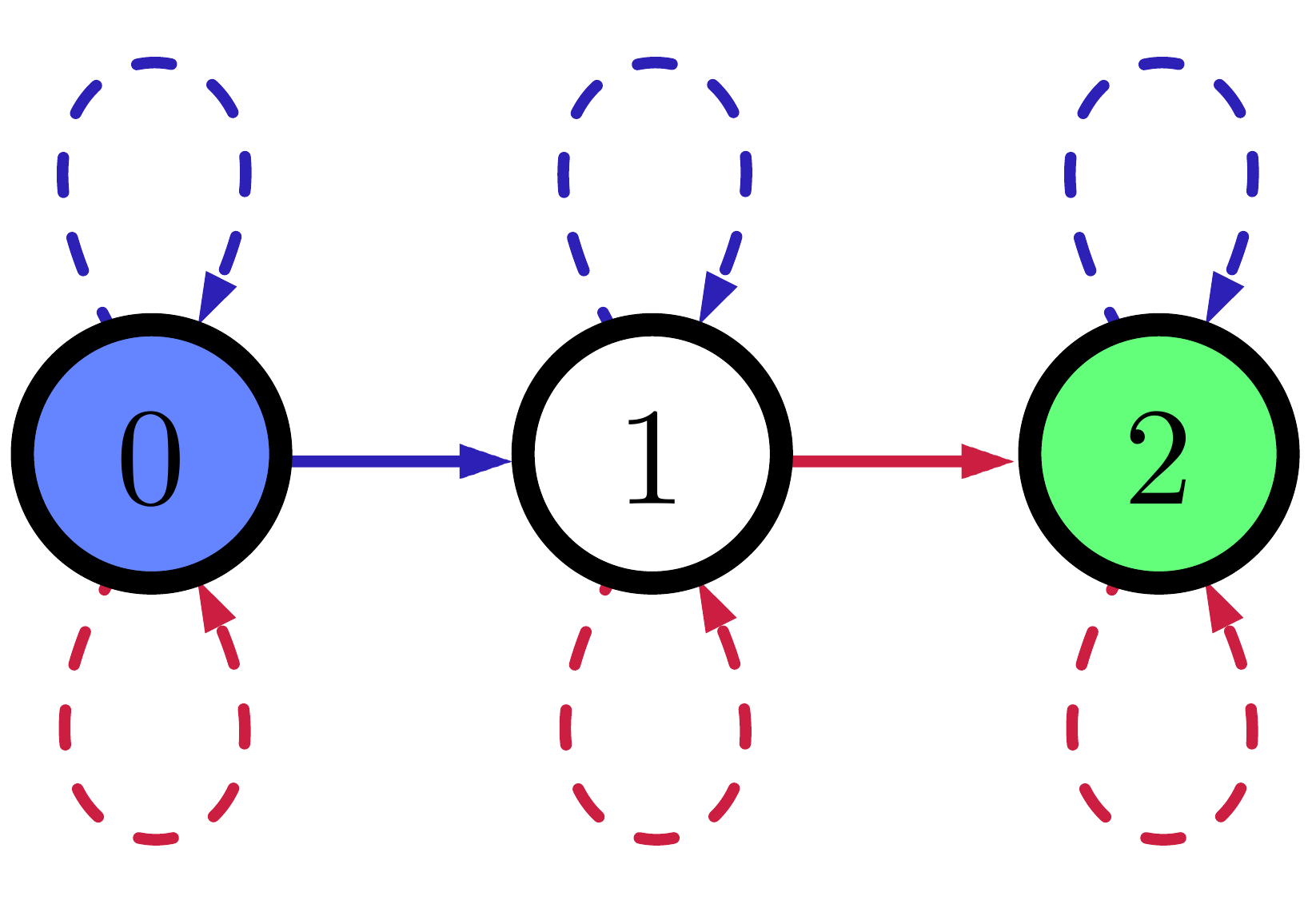}
\includegraphics[width=0.25\linewidth]{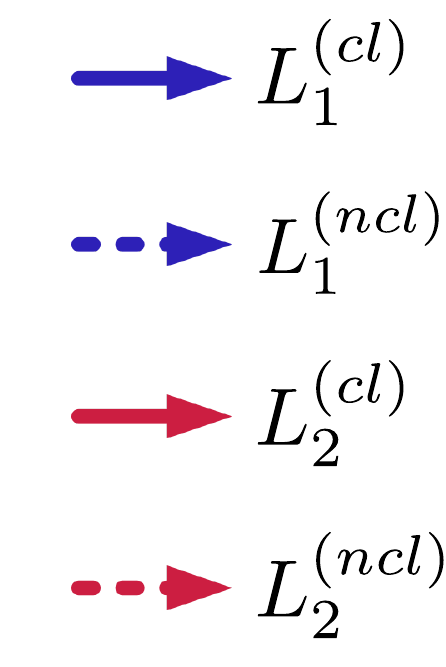}
\caption{\label{fig:basic_QSM} A basic example of the QSM multigraph, describing the available coupling family for a three-state system. The steering options are represented by the coupling operators $V_{1}=\gamma_1\ket{1}\bra{0}$ and $V_{2}=\gamma_2\ket{2}\bra{1}$.  The starting state is $0$, marked in blue, and the target state is $2$, marked in green. The optimal coordination policy of the two steering operations is straightforward: one needs to first repeatedly apply the $V_1$-steering until a click is obtained, and then the $V_2$-steering until a click is obtained. Compared to the passive steering which iterates between $V_1$ and $V_2$ regardless of measurement outcomes, this directly yields a 2-fold speedup in the average performance.}
\end{figure}

The steering graphs introduced above can now be used to create a Quantum State Machine. For $n_V$ elements $V_s(\textbf{p})$ in the available coupling family, there exist $2 n_V$ graphs corresponding to steering maps $\Lambda^\text{(cl, ncl)}_{V_{s}(\bf{p})}$, because of the two possible measurement outcomes for each of the couplings. The QSM for the steering protocol is then obtained as a collection of these graphs. It can be represented as a colored multigraph, where each steering graph is represented as a single-color subgraph (Fig.~\ref{fig:basic_QSM}). Consequently, in a QSM multigraph there may be multiple edges going from any vertex $\alpha$ into any other vertex $\beta$ (making it a multigraph rather than a simple graph), but at most one such edge for each color. 

Let us now consider our original task of finding the accelerated navigation protocol. To make use of the QSM construction in this context, we will restrict our consideration to bases $\{\ket{\phi_\beta}\}$ where one of the basis states is the target state $\ket{\psi_{\mathrm{targ}}}$ itself. In such a case, state $\ket{\psi_{\mathrm{targ}}}$ corresponds to a marked vertex in the graph, and the goal of the steering protocol becomes to drive the system state to that vertex. The goal of optimizing this protocol may then look similar to a known problem of finding the shortest path to the marked vertex on a weighted graph. This problem is standard in graph theory and can be solved as such. Can such a solution be used to design the navigation protocol?

As we will see in Sec.~\ref{sec:QSM_quantum_elements}, this analogy is not complete, since the quantum evolution on the graph goes beyond the classical path-on-the-graph picture. 
This aspect creates an obstacle to directly applying the graph exploration algorithms to facilitate our protocol speed-up. Fortunately, this difficulty can be properly accounted for in some cases, as we will see in Sec.~\ref{sec:QSM_coarse-graining}. In those cases, the ``semi-classical heuristics'' of graph exploration may indeed be applied. Finally, in Sec.~\ref{sec:QSM_W_preparation}, we will apply this approach to actively prepare the $W$-state, with a factor $f=3.5$ improvement compared to the passive protocol.

\subsection{Quantum subgraphs in a QSM}
\label{sec:QSM_quantum_elements}

Let us now compare our QSM navigation task to the standard problem of graph exploration. Our goal is to identify the differences between the two, which prevent us from applying the graph exploration techniques directly to QSM navigation. First of all, the state of the system in graph exploration is at all times represented by a single vertex.
The system in a QSM, on the other hand, is generally represented by a superposition over multiple vertices. Furthermore, in graph exploration, the state is modified by following one of the edges. A steering action in a QSM, in contrast, corresponds to a whole collection of edges -- i.e., a single-colored steering graph in the QSM multigraph.

\begin{figure}[t]
\includegraphics[width=0.45 \linewidth]{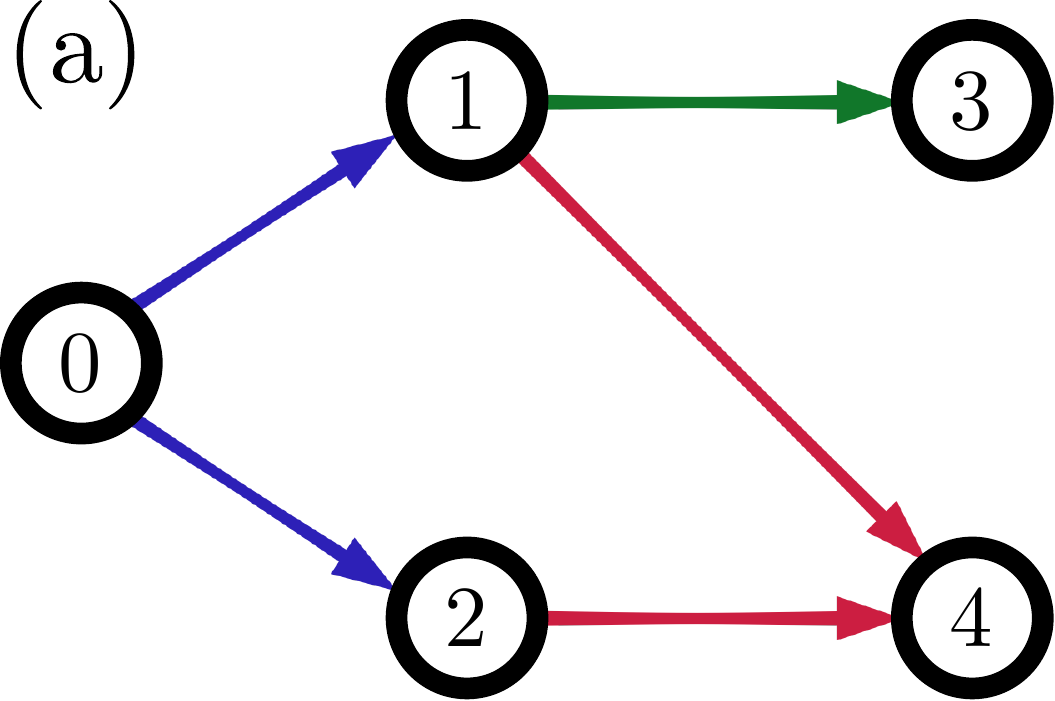}
\includegraphics[width=0.45 \linewidth]{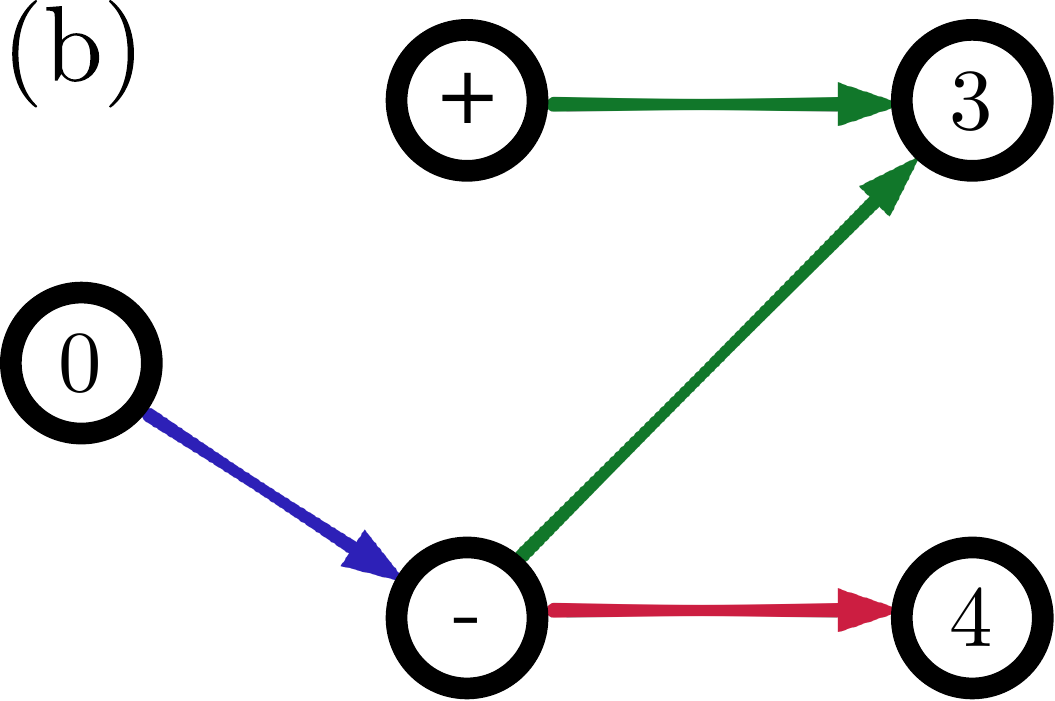}
\includegraphics[width=0.95 \linewidth]{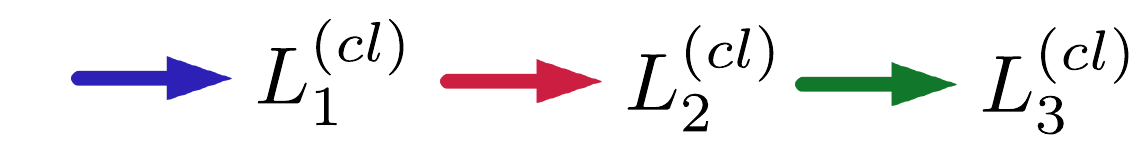}
\caption{\label{fig:quantum_QSM} Possible configurations of quantum subgraphs in a QSM, exemplified by the 5-vertex subgraph of a hypothetical QSM. (a) The click-action graphs for the three operators $V_{1,2,3}$ that form the family of couplings. The operators have the form $V_1=\gamma_1(\ket{1}-\ket{2})\bra{0}$, $V_2=\gamma_2\ket{4}(\bra{1}-\bra{2})$, $V_3=\gamma_3\ket{3}\bra{1}$. The graphs for the no-click actions are not shown, as their form can be deduced from the graphs for click actions. In the present basis, the $V_1$-click is manifest as a superposition, the $V_2$-click -- as an interference, and the $V_3$-click corresponds to a semiclassical evolution. (b) Quantum State Machine for the coupling family from the previous panel, depicted in a different basis. The basis transformation is $\ket{\pm}=(\ket{1}\pm\ket{2})/\sqrt{2}$. In this case, the basis transformation removes the quantum elements in the $L^{(cl)}_{1,2}$ graphs, however, it turns $L^{(cl)}_{3}$ into an interference element. Note that there is no basis transformation that would turn such a QSM into a classical one. This statement follows from the uniqueness of the Jordan canonical form for operators $V_2$ and $V_3$.}
\end{figure}

Some steering graphs may induce quantum effects, such as superposition and interference. For instance, the steering action whose graph contains two outgoing edges from a given vertex (e.g., vertex $0$ for graph $L^\text{(cl)}_1$ in Fig.~\ref{fig:quantum_QSM}a), can create a nontrivial quantum superposition. If a state is given by a superposition of multiple vertex states, it may further undergo quantum interference. In particular, this can be facilitated by a steering action whose graph contains a vertex with two incoming edges (e.g., vertex $4$ for graph $L^\text{(cl)}_2$ in Fig.~\ref{fig:quantum_QSM}a). 
In general, a notion of ``superposition subgraphs'' and ``interference subgraphs'' of a steering graph can be defined:
\begin{enumerate}
    \item Superposition subgraph is a subgraph of a steering graph span by multiple (more than one) edges outgoing from a single vertex.
    \item Interference subgraph is a subgraph of a steering graph span by multiple edges incoming to a single vertex. 
\end{enumerate}
Collectively, we will refer to such interference and superposition subgraphs of a single steering graph as its quantum subgraphs. If the quantum subgraphs are absent in the QSM, we will refer to it as a classical QSM. In other words, in a classical QSM, each vertex has at most one outgoing and at most one incoming edge of any given color. 

If a QSM is classical, optimization of the navigation protocol can essentially be reduced to classical graph exploration.
For a simple example of a classical QSM and the way to optimize the respective state preparation, consider the 3-level steering actions described in Fig.~\ref{fig:basic_QSM}. 
Note that optimization of the classical QSM also applies to the case when the starting state is a superposition of multiple vertex states. If the steering operations contain no quantum subgraphs, the quantum superposition is equivalent to a probabilistic mixture for the sake of the protocol optimization, and the optimal navigation pattern can be extracted accordingly. 

As the form of the steering graph depends on the choice of basis, it is conceivable that the number of quantum subgraphs in such a graph in some cases can be reduced by changing the basis (compare Fig.~\ref{fig:quantum_QSM}a and b). However, using a change of basis to remove all the quantum subgraphs in an arbitrary QSM is generally impossible (see Fig.~\ref{fig:quantum_QSM}).

\subsection{Coarse-grained QSM. Semiclassical heuristic for navigation}
\label{sec:QSM_coarse-graining}

We now focus on the steering protocols whose QSM cannot be made classical via a basis transformation. In such a case, it may still be possible to optimize it via a classical graph exploration heuristic. For that, we propose to coarse-grain the
QSM by grouping subsets of its vertices into single block-vertices. The coarse-grained QSM would consist of graphs drawn between such block-vertices. The block-vertex containing the target vertex can be considered as the target block-vertex.

\begin{figure}[t]
\includegraphics[width=0.9\linewidth]{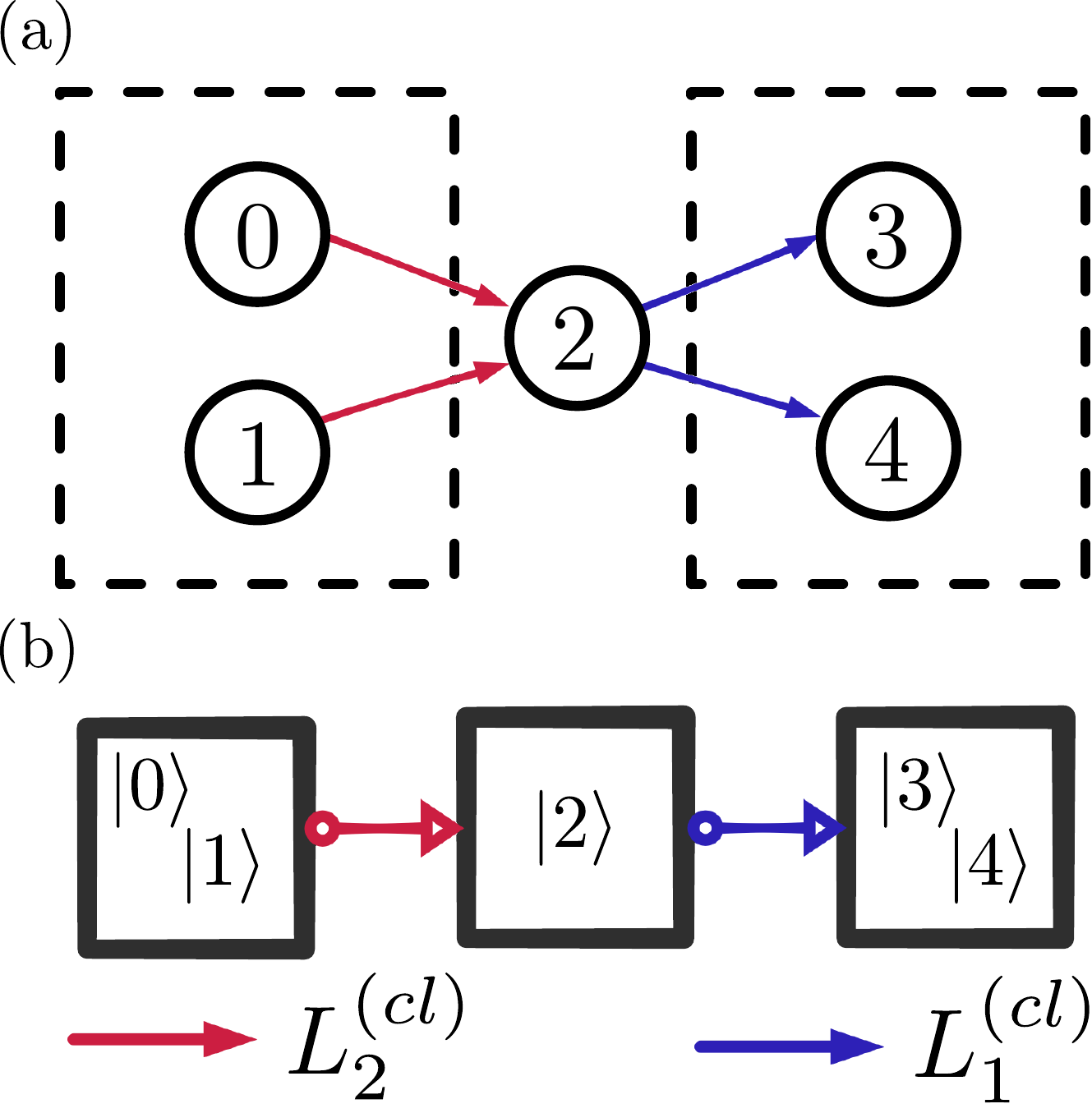}
\caption{\label{fig:QSM_coarse} Semiclassical coarse-graining applied to a QSM. (a) A 5-state part of a QSM with two quantum subgraphs: interference subgraph realized by $L^{(cl)}_2$ and a superposition subgraph realized by $L^{(cl)}_1$. Since pairs of states $\{\ket{0},~\ket{1}\}$ and $\{\ket{3},~\ket{4}\}$ fall under conditions described in Sec.~\ref{sec:QSM_coarse-graining}, these are to be grouped together in a coarse-grained QSM. (b) Simplified depiction of a coarse-grained QSM, obtained from (a). }
\end{figure}

An inter-block edge between two block-vertices is drawn, if the original QSM has at least one edge connecting the vertices inside the respective block-vertices. For the coarse-graining to be useful for our purposes, it should be done in such a way that all of the resulting QSM graphs have a classical structure. Namely, the coarse-grained graph should not have quantum subgraphs, e.g. realizing superposition or interference between the block-vertices (in analogy to Sec.~\ref{sec:QSM_quantum_elements}). To satisfy this requirement, the following rule for vertex grouping can be employed (cf.  Fig.~\ref{fig:QSM_coarse}):
\textit{if two edges of the same color are simultaneously coming in or out of a given vertex, the two vertices at the other ends of these edges should be grouped within one effective block-vertex.}
This rule manifestly yields basis-dependent groupings, since the very presence of quantum subgraphs in a QSM is basis-dependent. Thus, a smart choice of the basis may allow for an efficient and simpler coarse-grained graph. 
Designing a general explicit algorithm for finding the optimum basis for an arbitrary QSM is a highly non-trivial task. Heuristically speaking, a convenient choice of the basis should be the one that results in the minimum number of quantum subgraphs in a QSM before coarse-graining.

\begin{figure}[t]
\includegraphics[width=0.9\linewidth]{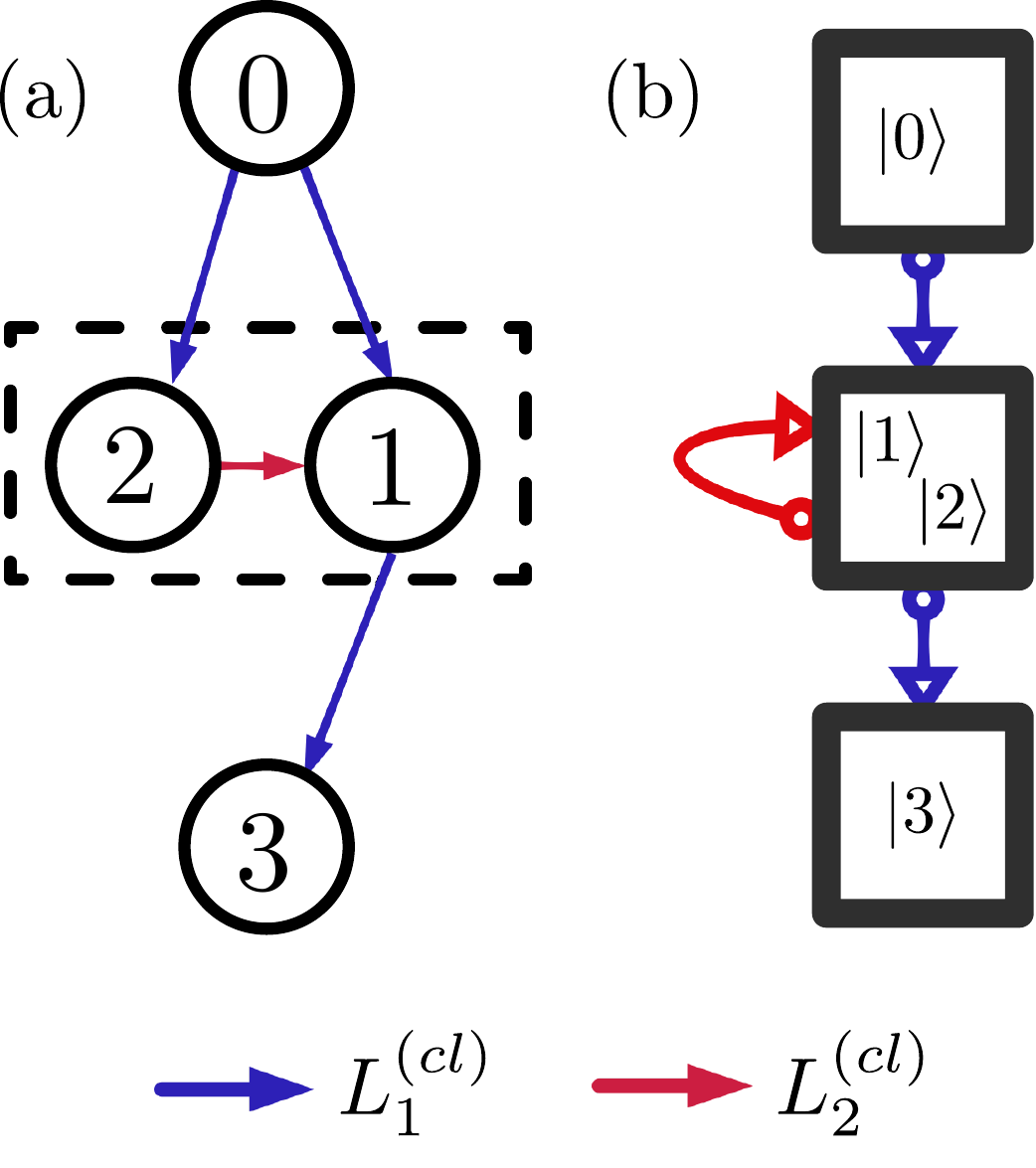}
\caption{\label{fig:QSM_ancillary} Illustration of ancillary couplings in the context of QSM coarse-graining (a) A 4-state part of a QSM that is subject to coarse-graining, featuring non-trivial actions by couplings denoted as $V_1$ and $V_2$. States $\ket{1}$ and $\ket{2}$ are to be grouped together since they are both targets in a superposition subgraph. (b) The coarse-grained version of the same QSM. The block $\{\ket{1}, \ket{2}\}$ is connected to state $\ket{3}$ through an outgoing edge of $L^\text{(cl)}_1$. However, from microscopic point of view exemplified in the first panel, no population can be transferred from state $\ket{2}$ to $\ket{3}$ unless the click action $\Lambda^\text{(cl)}_2$ is realized first. Therefore, including and applying $V_2$ as an ancillary coupling is required for a valid semiclassical coarse-graining of this QSM.}
\end{figure}

For the coarse-grained graph to be effectively classical, we desire to ignore details of the system evolution inside the subspace of a given block-vertex. Specifically, we aim to view every block-vertex as an effective single state of the system and assume that every edge allows transporting the system between such block-vertex states with no obstacles. If this was directly possible, and since the coarse-grained QSM by definition contains no quantum subgraphs, optimization of its exploration would have become a classical task. However, such an approximation scheme needs more careful justification. Every block fundamentally corresponds to a Hilbert subspace, and an inter-block edge is given by a $D_1\times D_2$ matrix of coefficients (where $D_1$ and $D_2$ are the internal dimensionalities of the linked blocks). Characterizing these effectively with single amplitudes may lead to erroneous navigation policies.
In particular, one state internal to a block-vertex might be untouched by an inter-block edge, i.e., it only yields zero matrix elements in a matrix characterizing the edge. If the edge is outgoing, a system initialized in the said state would not be able to escape the block-vertex using that edge alone (see Fig.~\ref{fig:QSM_ancillary}). 
This is in direct conflict with characterizing blocks and inter-block edges with single amplitudes.  
For an incoming edge, a similar problem may arise: some states inside a block-vertex might not get populated when that edge is activated. This may become detrimental for the navigation protocol based on a coarse-grained QSM, especially if the unavailable state in question is the final target of the protocol.

Such difficulties may be overcome, if some of the couplings given in a QSM allow for an internal mixing of the subspace (represented by a self-loop on the block-vertex in the respective $L^\text{(cl)}$-graph). Applying such a coupling in the protocol would allow one to make the block-vertex accessible to all the edges that are connected to it (see Fig.~\ref{fig:QSM_ancillary}), via a sufficient number of clicks. In the scenarios described above, where additional couplings are needed to turn a block-vertex into an effective single vertex, we will refer to such couplings as ancillary couplings. Note that given a coupling family, there is no guarantee that the ancillary couplings needed for the exploration of every block-vertex, are available. In this work, we restrict our further consideration to the coarse-grained QSMs, where the ancillary couplings happen to be present wherever needed. Every block-vertex can then be made accessible to the outgoing edges, and the target state is ensured to be reachable once the target block is reached. In this case, we consider the coarse-grained QSM as effectively semiclassical.

To design an active steering policy within the coarse-grained approach, we note that the navigation protocol has the following structure. The system state can be transported between block-vertices, and eventually steered to the target block-vertex. After that, either the target state is reached already (one can obtain this information from the simulated copy of the system), or it can be reached after applying ancillary couplings on the target block-vertex. The cost of the protocol can now be broken into two parts. The first is the cost of exploring the coarse-grained graph using the inter-vertex edges. The second is the dwell time inside the block-vertices, which is spent applying the ancillary couplings. If we could find the route through the graph that minimizes the combination of these two components, it would solve our optimization problem exactly. There is a problem, however: both the inter-vertex travel time and the block-vertex dwell time depend on the microscopic details of the evolution internal to the block-vertices. The coarse-grained geometrical information would therefore not suffice to \textit{exactly} derive the optimal policy. On the other hand, fully accounting for quantum-mechanical microscopics is prohibitively hard. Instead, we will use the semiclassical QSM to obtain an \textit{approximately} optimal active policy.

Let us assign every inter-block edge a characteristic traversal time, and every block-vertex a characteristic dwell time. For this, we use the matrices for click transitions between blocks $i$ and $j$ (the case of ancillary couplings given by $i=j$). Denote these as $L^\text{(cl)}_{i,\alpha;j,\beta}$, implying that only matrix elements with states from blocks $i$ and $j$ are included. In that case, the effective transition amplitude between blocks $i$ and $j$ can be defined as operator norm $L^\text{(cl)}_{i,j}=\|L^\text{(cl)}_{i,\alpha;j,\beta}\|$, and characteristic traversal (dwell if $i=j$) time $\Delta\tau_{i,j}=(L^\text{(cl)}_{i,j})^{-2}$. This reduces to the average traversal time for the case of a genuinely classical graph, with an amplitude $\gamma\delta t$ connecting two states implying duration of $\Delta\tau=(\gamma\delta t)^{-2}$ for traversal (cf. Sec.~\ref{sec:single_qubit}). 

With characteristic times $\Delta\tau_{i,j}$ assigned, the time-cost of following a specific path through this graph can be estimated as a combined characteristic time of all the edges and vertices crossed along the way. The desired path will be the one that optimizes this expected time. As previously discussed, this semiclassical calculation may not yield an exactly optimal navigation policy. However, in many practical cases such an active protocol should still be quicker compared to its completely passive version. One example of such a practical improvement is given below.

\subsection{W-state preparation}
\label{sec:QSM_W_preparation}

\begin{figure}[t]
\includegraphics[width=\linewidth]{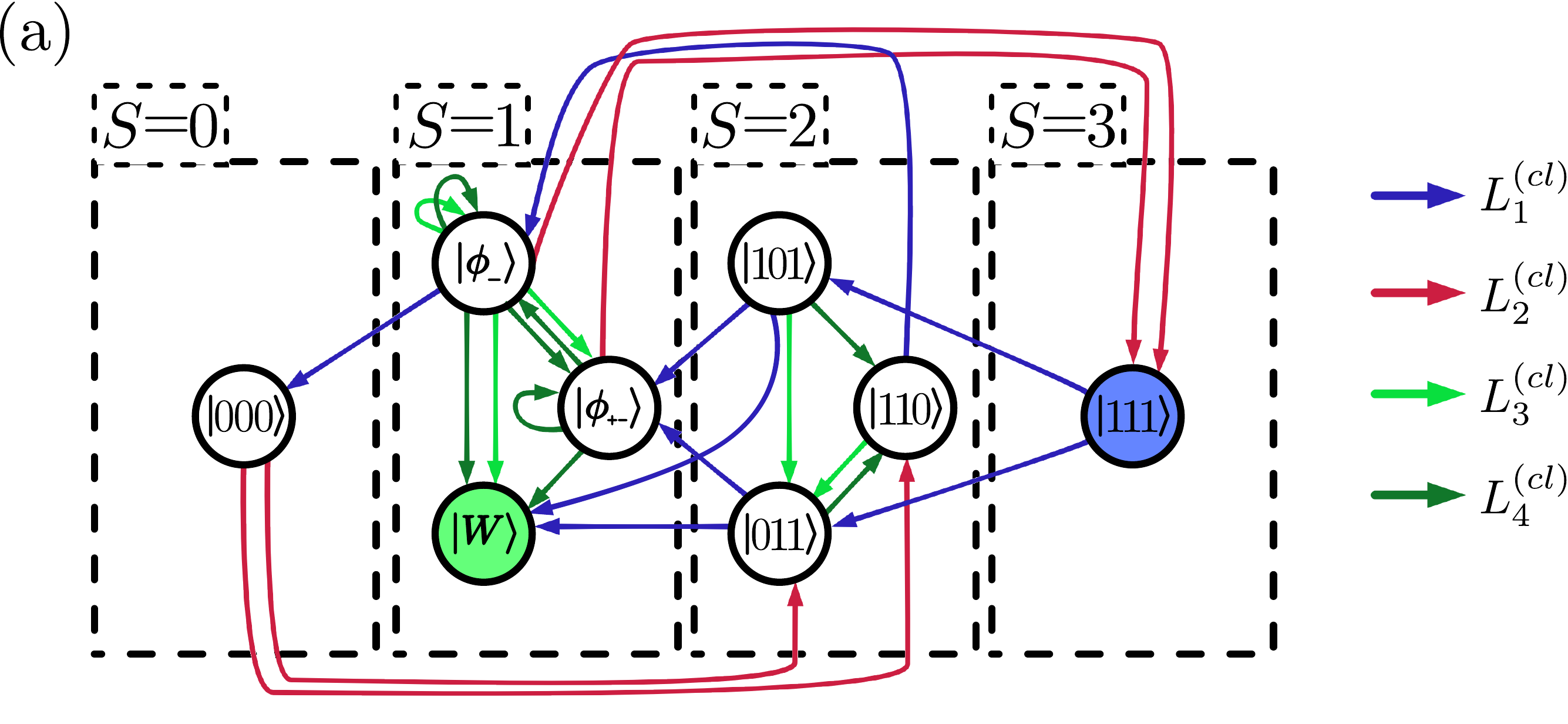}
\includegraphics[width=\linewidth]{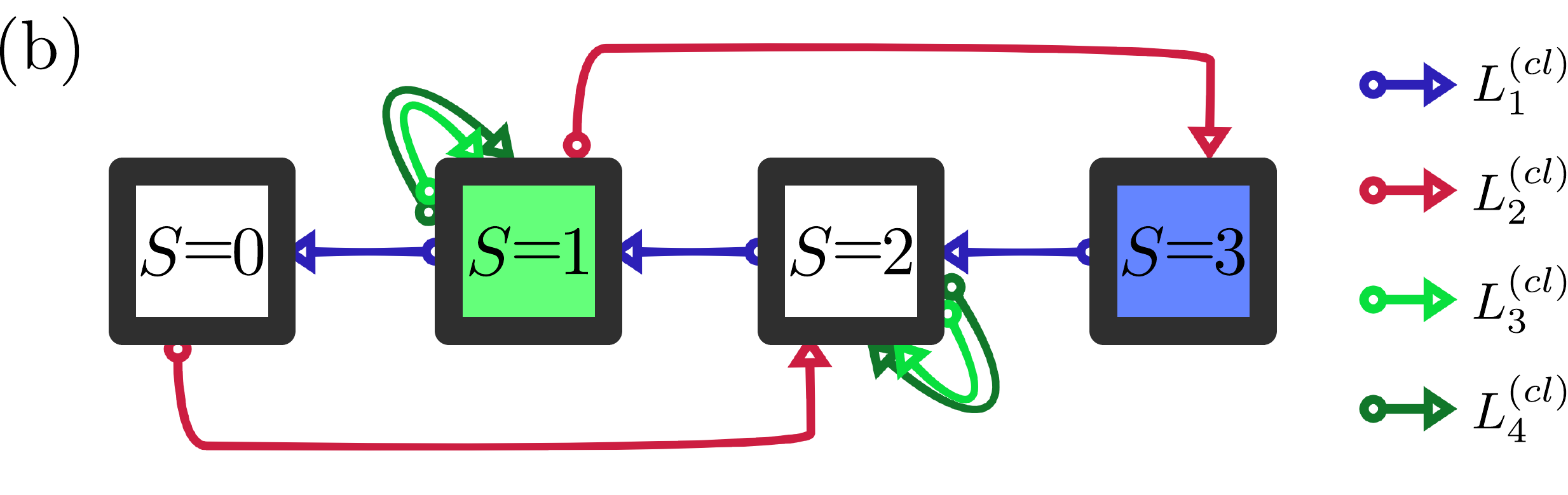}
\caption{\label{fig:W_QSM} 
Measurement-driven navigation towards the 3-qubit $W$-state: QSM representation. (a) Steering with couplings Eqs.~\eqref{eq:W_V_1}-\eqref{eq:W_V_4}. The vertices in the single-excitation subspace are given by states $\ket{W}$, $\ket{\phi_-}\equiv\frac{1}{\sqrt{2}}(\ket{100}-\ket{001}) $, and $\ket{\phi_{+-}}\equiv\frac{1}{\sqrt{6}}(\ket{100}-2\ket{010}+\ket{001})$. (b) The coarse-grained version of the above QSM. The vertices are labeled by the excitation number. From perspective of Sec.~\ref{sec:QSM_coarse-graining}, couplings $2$ and $3$ play the ancillary role. Indeed, those couplings mix the internal structure of the block-vertices, allowing one to eventually steer the state to the target $\ket{W}$.}
\end{figure}

To illustrate the principles of the QSM framework, we consider the coarse-graining approach to the navigation of a 3-qubit system from $\ket{111}$ to a so-called W-state \cite{dur2000three}. This state has the following form:
\begin{equation}
W = \frac{1}{\sqrt{3}} (\ket{100}+\ket{010}+\ket{001}).
\end{equation}
To define the measurement-based protocols, we choose the following $2$-local family of couplings (assuming labels A, B, and C for the qubits):
\begin{align}
V_1 & = \sigma^-_A-\sigma^-_B, \label{eq:W_V_1}\\
V_2 & = \sigma^+_A\sigma^+_B-\sigma^+_B\sigma^+_C, \label{eq:W_V_2}\\
V_3 & = \sigma^-_A\sigma^+_B - P^0_A P^1_B, \label{eq:W_V_3}\\
V_4 & = \sigma^+_B\sigma^-_C - P^1_B P^0_C. \label{eq:W_V_4}
\end{align}
Here, $\sigma^{\pm}=\frac{1}{2}\left(\sigma^x\pm i\sigma^y\right)$ and $P^a=\ket{a}\bra{a},\ a=0,1$.
A passive version of the protocol would amount to blindly alternating between the steering actions with different $V_i$. This does yield the target state if the steering is applied a sufficient number of times (Fig.~\ref{fig:W_QSM}.b). A cost-function based \textit{active} policy (Sec.~\ref{sec:quantum_compass}) can also be introduced, greedily choosing between $V_i$ based on the expected gain in the target state fidelity.

\begin{figure}[t]
\includegraphics[width=0.99\linewidth]{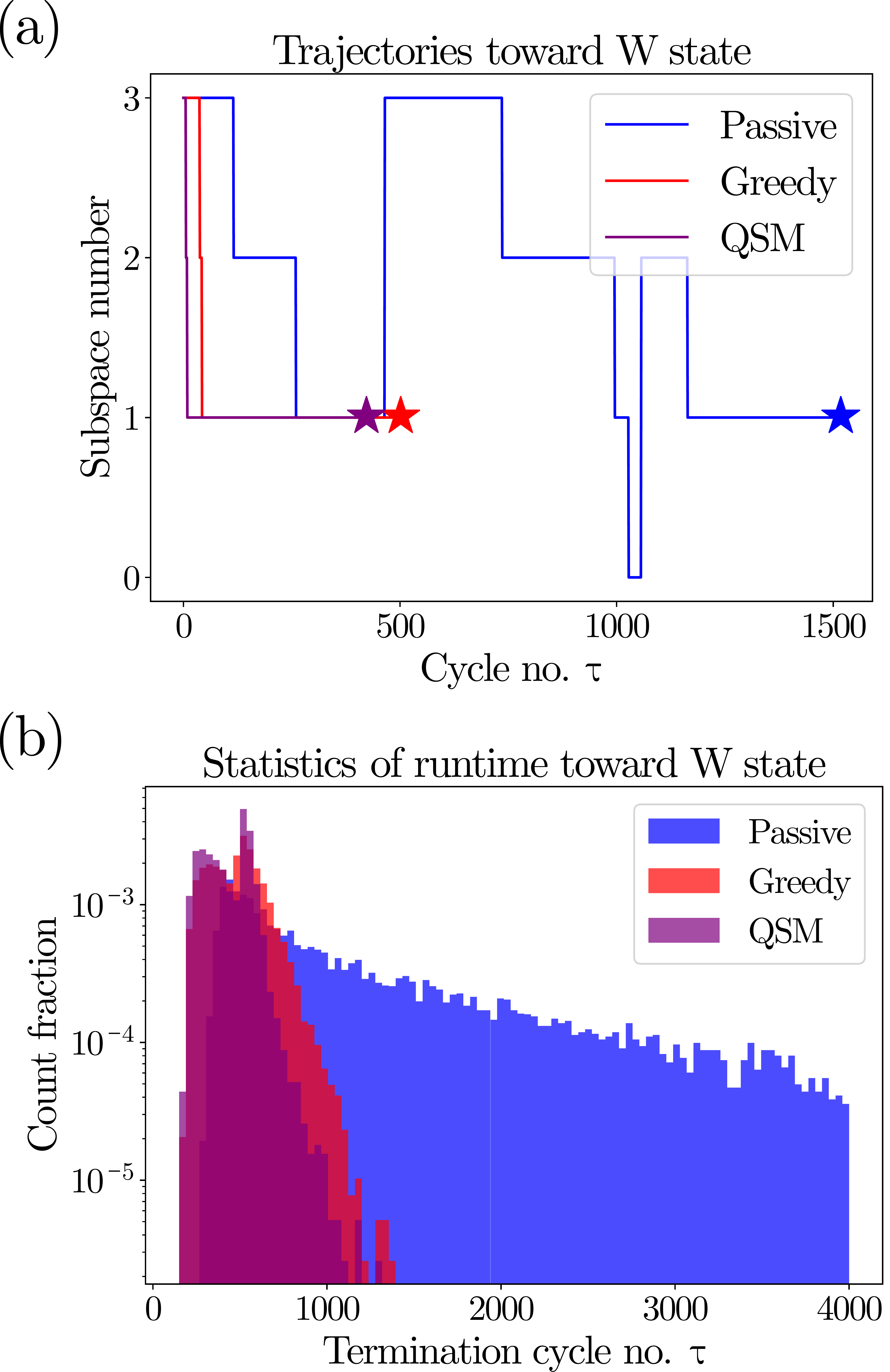}
\caption{\label{fig:performance_QSM} Performance of the QSM-based protocol compared to its passive and greedy counterparts. The protocols are simulated with $\delta t = 0.1$ and the target precision $\epsilon = 0.01$. (a) Typical trajectories of the passive and active protocols, across the excitation number sectors $S$ occupied by the running state. Displayed are trajectories that yield the runtimes approximately equal to average runtimes $\tau^\mathrm{(QSM)}_\mathrm{av}\simeq430$, $\tau^\mathrm{(pas)}_\mathrm{av}\simeq1500$, and $\tau^\mathrm{(greedy)}_\mathrm{av}\simeq490$. Passive protocol switches between different $S$ multiple times before eventually reaching the target. The greedy protocol behaves similarly to QSM-based one, but stays longer in the $S=3$ state, accounting for its relative slowdown. (b) The logarithmic histogram over the protocol runtimes. It can be seen that the QSM-based protocol consistently outperforms the passive protocol and often outperforms the greedy one. Each histogram was obtained from $10^4$ numerical simulations, and truncated at $\tau=4000$ for better presentation.}
\end{figure}

To design a QSM-based active policy, consider a multigraph representation of the coupling family.
It is shown in Fig.~\ref{fig:W_QSM}a. Note that this QSM has multiple quantum subgraphs. Therefore, to employ a feedback policy, it should be subjected to the coarse-graining technique as outlined in Sec.~\ref{sec:QSM_coarse-graining}. It proves useful to coarse-grain the Hilbert space by the total excitation number $S\in\{0,1,2,3\}$ -- this results in a semiclassical QSM, as desired (Fig.~\ref{fig:W_QSM}b). Given the all-up starting state of the evolution, one designs an active policy that leads to the target state in a classically optimal way:
\begin{enumerate}
    \item Repeat $V_1$-steering until a click is obtained;
    \item Repeat $V_1$-steering until another click is obtained;
    \item Alternate $V_3$- and $V_4$-steering until the target state is reached (with fidelity error below $\epsilon$).
\end{enumerate}
This protocol moves the state of the system from the triple excitation state to the double-excitation subspace (stage 1), then to single-excitation subspace (stage 2), and then takes the system to the W-state in that subspace (part 3). 

The performance of the above QSM-based active protocol can be numerically compared to its passive and greedy counterparts (Fig.~\ref{fig:performance_QSM}). 
The average runtimes of these are, respectively, $\tau^\mathrm{(QSM)}_\mathrm{av}\simeq430$, $\tau^\mathrm{(pas)}_\mathrm{av}\simeq1500$, and $\tau^\mathrm{(greedy)}_\mathrm{av}\simeq490$. To understand the reason for slower performance of the passive protocol, note that it can move the system in and out of one excitation subspace before the target state is reached (Fig.~\ref{fig:performance_QSM}a). The greedy cost-function based policy avoids this issue and thus offers a speedup $f_{\mathrm{greedy}}\simeq 3.1$, however, it underperforms compared to QSM-based policy ($f_{\mathrm{QSM}}=3.5$). The reason is that it suffers from the landscape flatness (Sec.~\ref{sec:orthogonality_catastrophe}) issue. Before the system moves from $S=3$ to $S=2$, no coupling action is capable of directly achieving nonzero target state fidelity. The resulting delay can be seen in the typical protocol trajectories, see Fig.~\ref{fig:performance_QSM}a. This disparity between $f_{\mathrm{greedy}}$ and $f_{\mathrm{QSM}}$ highlights the complementary nature of the two navigation approaches presented in this work. To apply measurement-driven navigation in the best way possible, one has to identify the better approach based on the target state and the coupling operators available.

\section{Discussion and conclusions \label{sec:discussion}}

In this work, we have put forward the concept of measurement-driven active-decision steering of quantum states. We have developed steering protocols in which the measurement readouts are used to adjust the measurement protocol on-the-go, yielding significant acceleration of state preparation relative to passive steering. The possibility of exploiting the readouts explored here is the great advantage of measurement-based steering over drive-and-dissipation state preparation (which is largely equivalent to ``blind'' steering). While our approach has sweeping applicability, here we have chosen to focus on active measurement-driven steering as applied to the most challenging case of many-body quantum systems with entangled target states.

To satisfy physical (locality) constraints on system-detector couplings, we have proposed a scheme, based on parent Hamiltonian construction, for identifying feasible couplings. Employing such couplings, we have developed and analyzed Hilbert-space-orientation techniques for measurement-driven steering. A central ingredient here has been to develop feedback policies based on detector readouts. 

The first Hilbert-space path-finding technique is based on a cost function, evaluating the running fidelity to the target state. We have shown a substantial (up to $23$-fold) speedup of steering, employing this approach for preparation of the ground state of the AKLT model. For randomly generated MPS targets, the speedup from this method ranges at least up to $f_\mathrm{max}\sim 10^3$, with an average value of $f_\mathrm{av}\sim 10^1$ ($N=5$ spin-1 system). Intriguingly, our numerics strongly suggests the growth of speedup with system size for MPS targets whose parent Hamiltonians have a significant spectral gap. 

A second method comprises mapping out the available measurement actions onto a Quantum State Machine (QSM), using a coarse-grained version of the corresponding graphs in Hilbert space. This approach is of conceptual significance complementary to the greedy method, being distinct in its principle and potentially surpassing the performance of greedy protocol for certain targets. We have shown an example of $W$-state preparation, where a QSM-based method provides a speedup $f_\mathrm{QSM}=3.5$ that is higher compared to the greedy approach ($f_\mathrm{greedy}=3.1$).

While we have limited ourselves here to specific examples, our schemes are of general applicability. They open the door to the design of efficient and high-quality state engineering, adiabatic state manipulation, and, possibly, quantum information processing. Moreover, steering protocols are subject to errors, both ``static'' (choice of steering parameters) and ``dynamics'' (noise) \cite{Edd}, in addition to a reduced ``detection efficiency'' discussed in Sec.~\ref{failed-active}. Active decision-making steering may be designed to reduce the effect of such errors, by including self-corrections based on the recorded sequence of readouts. Importantly, compared to the greedy protocols, the effect of ``measurement imperfections'' is expected to be reduced for QSM strategies, as these operate with coarse-grained objects at the semiclassical level.  

One may envision a host of further directions to generalize and develop the ideas of active steering. For example, the greedy minimization  of our cost function may be further improved by finding other metrics of local ``steepest decent.'' Further, one may systematically investigate less local (less greedy) optimization of the cost function, e.g., looking multiple cycles ahead. 
Another potential advantage of our protocols relies on the following observation: in the context of passive steering, one imposes constraints concerning locality (e.g., how many spins can be coupled to a local detector), and certain types of coupling terms. Given such constraints, not all target states are reachable. The introduction of active steering may overcome this handicap of target-state accessibility.  

One may also combine the dynamics incorporated here with the inherent unitary evolution of the system at hand (due to a system-only Hamiltonian). Consider the context of passive (blind) measurement-induced steering, which, in the continuum time limit, leads to Lindbladian dynamics. Then, the addition of Hamiltonian dynamics enriches the variability of steering, allowing, for example, to obtain mixed states by design \cite{kumar2020engineering}. It is intriguing to investigate how the addition of Hamiltonian dynamics extends or improves active steering, thus marrying the frameworks
of closed-loop quantum control for Hamiltonian-based state preparation and active-decision measurement-based steering. We expect, in particular, that active-decision strategies would allow one to steer the system to a pure target state even in those cases where the passive protocols yield mixed states. Systematic study of the combined action of active-decision measurement protocols and system-only Hamiltonian is an extremely interesting challenge for the future.

Further extensions of our approach include applications of QSM protocols to larger and more complex systems, going beyond a three-qubit setup. Optimizing such protocols may involve automatization of the creation and analysis of QSMs, e.g., for finding an optimal basis automatically,
in similarity with quantum annealing, but now at the level of measurement operators. One may foresee a protocol, where combining local rotation of the basis states with a renormalization-group procedure, a structure of “quantum vertices” that are interconnected semiclassically emerges. This would then, in particular, admit a QSM engineering of MPS targets. 

Finally, one may envision using machine learning to find more optimized navigation protocols (see \cite{borah2021measurement, bondarenko19quantum, dehg22} for related work). Given the delayed-reward setting at hand, a reinforcement learning strategy such as Q-learning \cite{watkins1989learning} or SARSA \cite{rummery1994online} might be the most appropriate choice.

\section*{Acknowledgments}

We thank D. Bondarenko,
X.~Bonet-Monroig,
J.~Chalker, R.~Egger,  C.~Koch, P.~Kumar, E.~Medina~Guerra, S.~Morales, G.~Morigi,
S.~Polla,
R.~Raussendorf,
S.~Roy, K.~Snizhko,  and A.~Zazunov for discussions. 
The work was supported by the Deutsche Forschungsgemeinschaft (DFG): Project No. 277101999 -- TRR 183 (Project C01) and Grants No. EG 96/13-1 and No. GO 1405/6-1, as well as by the Israel Science Foundation and the Helmholtz International Fellow Award, the National Science Foundation through
award DMR-2037654, the US-Israel Binational 
Science Foundation (BSF), and the Netherlands
Organization for Scientific Research (NWO/OCW).


\begin{thebibliography}{99}
\bibitem{footnote1} This contrasts with the common definition of frustration in Hamiltonians, by which the locally projective $H_{\mathrm{par}}(\psi_{\mathrm{targ}})$ is frustration-free despite the non-commutation of its terms.
\bibitem{footnote2} In this section, we consider the all-down product state as our starting state. In both passive and active protocol, we assume each coupling to be applied multiple times until one either receives a click, or no-click for an asymptotically long time. Such a repeated application of a single coupling is then counted as a single protocol cycle. We take this approach for a practical purpose because simulating such protocols is more accessible numerically.
\end{thebibliography}
\end{document}